\documentclass[twocolumn,journal]{IEEEtran}
\usepackage[T1]{fontenc}
\usepackage[latin9]{inputenc}
\usepackage{prettyref}
\usepackage{float}
\usepackage{mathtools}
\usepackage{amsmath}
\usepackage{amsthm}
\usepackage{amssymb}
\usepackage{graphicx}
\usepackage[unicode=true,
 bookmarks=true,bookmarksnumbered=true,bookmarksopen=true,bookmarksopenlevel=1,
 breaklinks=false,pdfborder={0 0 1},backref=false,colorlinks=false]
 {hyperref}
\hypersetup{pdftitle={Your Title},
 pdfauthor={Your Name},
 pdfborderstyle=,pdfpagelayout=OneColumn,pdfnewwindow=true,pdfstartview=XYZ,plainpages=false}

\makeatletter

\providecommand{\tabularnewline}{\\}
\floatstyle{ruled}
\newfloat{algorithm}{tbp}{loa}
\providecommand{\algorithmname}{Algorithm}
\floatname{algorithm}{\protect\algorithmname}

\theoremstyle{plain}
\newtheorem{thm}{\protect\theoremname}
\theoremstyle{plain}
\newtheorem{lem}{\protect\lemmaname}

\usepackage{algorithmic}

\usepackage{cite}

\providecommand{\lemmaname}{Lemma}
\providecommand{\theoremname}{Theorem}

\makeatother

\providecommand{\lemmaname}{Lemma}
\providecommand{\theoremname}{Theorem}

\begin{document}
\title{Parameter Estimation of Heavy-Tailed AR Model with Missing Data via
Stochastic EM}
\author{Junyan~Liu,~ Sandeep~Kumar,~and~Daniel~P.~Palomar,~\IEEEmembership{Fellow,~IEEE}\thanks{This work was supported by the Hong Kong RGC 16208917 research grant.}\thanks{The authors are with the Hong Kong University of Science and Technology,
Hong Kong (e-mail: jliubl@connect.ust.hk; eesandeep@ust.hk; palomar@ust.hk).}}
\maketitle
\begin{abstract}
The autoregressive (AR) model is a widely used model to understand
time series data. Traditionally, the innovation noise of the AR is
modeled as Gaussian. However, many time series applications, for example,
financial time series data, are non-Gaussian, therefore, the AR model
with more general heavy-tailed innovations is preferred. Another issue
that frequently occurs in time series is missing values, due to system
data record failure or unexpected data loss. Although there are numerous
works about Gaussian AR time series with missing values, as far as
we know, there does not exist any work addressing the issue of missing
data for the heavy-tailed AR model. In this paper, we consider this
issue for the first time, and propose an efficient framework for parameter
estimation from incomplete heavy-tailed time series based on a stochastic
approximation expectation maximization (SAEM) coupled with a Markov
Chain Monte Carlo (MCMC) procedure. The proposed algorithm is computationally
cheap and easy to implement. The convergence of the proposed algorithm
to a stationary point of the observed data likelihood is rigorously
proved. Extensive simulations and real datasets analyses demonstrate
the efficacy of the proposed framework. 
\end{abstract}

\begin{IEEEkeywords}
AR model, heavy-tail, missing values, SAEM, Markov chain Monte Carlo,
convergence analysis 
\end{IEEEkeywords}

\IEEEpeerreviewmaketitle{}

\vspace{-3mm}

\section{Introduction\label{sec:Introduction}}

\IEEEPARstart{I}{n} the recent era of data deluge, many applications
collect and process time series data for inference, learning, parameter
estimation, and decision making. The autoregressive (AR) model is
a commonly used model to analyze time series data, where observations
taken closely in time are statistically dependent on others. In an
AR time series, each sample is a linear combination of some previous
observations with a stochastic innovation. An AR model of order $p$,
AR($p$), is defined as

\vspace{-3mm}
 
\begin{equation}
y_{t}=\varphi_{0}+\sum_{i=1}^{p}\varphi_{i}y_{t-i}+\varepsilon_{t},\label{eq:ar model}
\end{equation}
where $y_{t}$ is the $t$-th observation, $\varphi_{0}$ is a constant,
$\varphi_{i}$'s are autoregressive coefficients, and $\varepsilon_{t}$
is the innovation associated with the $t$-th observation. The AR
model has been successfully used in many real-world applications such
as DNA microarray data analysis \cite{choong2009autoregressive},
EEG signal modeling\cite{schlogl2006analyzing}, financial time series
analysis\cite{tsay2005analysis}, and animal population study \cite{anderson2017black},
to name but a few.

Traditionally, the innovation $\varepsilon_{t}$ of the AR model is
assumed to be Gaussian distributed, which, as a result of the linearity
of the AR model, means that the observations are also Gaussian distributed.
However, there are situations arising in applications of signal processing
and financial markets where the time series are non-Gaussian and heavy-tailed,
either due to intrinsic data generation mechanism or existence of
outliers. Some examples are, stock returns \cite{tsay2005analysis,rachev2003handbook},
brain fMRI \cite{alexander2002detection,han2018eca}, and black-swan
events in animal population \cite{anderson2017black}. For these cases,
one may seek an AR model with innovations following a heavy-tailed
distribution such as the Student's $t$-distribution. The Student's
$t$-distribution is one of the most commonly used heavy-tailed distributions
\cite{lange1989robust}. The authors of \cite{tiku2000time} and \cite{tarami2003multi}
have considered an AR model with innovations following a Student's
$t$-distribution with a known number of degrees of freedom, whereas
\cite{nduka2018based} and \cite{christmas2011robust} investigated
the case with an unknown number of degrees of freedom. The Student's
$t$ AR model performs well for heavy-tailed AR time series and can
provide robust reliable estimates of the regressive coefficients when
outliers occur.

Another issue that frequently occurs in practice is missing values
during data observation or recording process. There are various reasons
that can lead to missing values: values may not be measured, values
may be measured but get lost, or values may be measured but are considered
unusable \cite{little2002statistical}. Some real-world cases are:
some stocks may suffer a lack of liquidity resulting in no transaction
and hence no price recorded, observation devices like sensors may
break down during measurement, and weather or other conditions disturb
sample taking schemes. Therefore, investigation of AR time series
with missing values is significant. Although there are numerous works
considering Gaussian AR time series with missing values \cite{dicesare2006imputation,ding2010time,Kharin2011,sargan1974missing},
less attention has been paid to heavy-tailed AR time series with missing
values, since parameter estimation in such a case is complicated due
to the intractable problem formulation. The frameworks for parameter
estimation for heavy-tailed AR time series in \cite{tiku2000time,tarami2003multi,nduka2018based,christmas2011robust}
require complete data, and thereby, are not suited for scenarios with
missing data. The objective of the current paper is to deal with this
challenge and develop an efficient framework for parameter estimation
from incomplete data under the heavy-tailed time series model via
the expectation-maximization (EM) type algorithm.

The EM algorithm is a widely used iterative method to obtain the maximum
likelihood (ML) estimates of parameters when there are missing values
or unobserved latent variables. In each iteration, the EM algorithm
maximizes the conditional expectation of the complete data likelihood
to update the estimates. Many variants of the EM algorithm have been
proposed to deal with specific challenges in different missing value
problems. For example, to tackle the problem posed by the intractability
of the conditional expectation of the complete data log-likelihood,
a stochastic variant of the EM algorithm, which approximates the expectation
by drawing samples of the latent variables from the conditional distribution,
has been proposed in \cite{delyon1999convergence,kuhn2004coupling}.
The stochastic EM has also been quite popular to curb the curse of
dimensionality \cite{nielsen2000stochastic,dicesare2006imputation},
since its computation complexity is lower than the EM algorithm. The
expectation conditional maximization (ECM) algorithm has been suggested
to deal with the unavailability of the closed-form maximizer of the
expected complete data log-likelihood \cite{meng1993maximum}. The
regularized EM algorithm has been used to enforce certain structures
in parameter estimates like sparsity, low-rank, and network structure
\cite{yi2015regularized}.

In this paper, we develop a provably convergent low cost algorithmic
framework for parameter estimation of the AR time series model with
heavy-tailed innovations from incomplete time series. As far as we
know, there does not exist any convergent algorithmic framework for
such problem. Following \cite{tiku2000time,tarami2003multi,nduka2018based},
here we consider the AR model with the Student's $t$ distributed
innovations. We formulate an ML estimation problem and develop an
efficient algorithm to obtain the ML estimates of the parameters based
on the stochastic EM framework. To tackle the complexity of the conditional
distribution of latent variables, we propose a Gibbs sampling scheme
to generate samples. Instead of directly sampling from the complicated
conditional distribution, the proposed algorithm just need to sample
from Gaussian distributions and gamma distributions alternatively.
The convergence of the proposed algorithm to a stationary point is
established. Simulations on real data and synthetic data show that
the proposed framework can provide accurate estimation of parameters
for incomplete time series, and is also robust against possible outliers.
Although here we only focus on the Student's $t$ distributed innovation,
the idea of the proposed approach and the algorithm can also be extended
to the AR model with other heavy-tailed distributions.

This paper is organized as follows. The problem formulation is provided
in \prettyref{sec:Problem-Formulation}. The review of the EM and
its stochastic variants is presented in \prettyref{sec:EM-and-Its-Stochastic-Variants}.
The proposed algorithm is derived in \prettyref{sec:SAEM-MCMC-for-Student's}.
The convergence analysis is carried out in \prettyref{sec:Convergence}.
Finally, Simulation results for the proposed algorithm applied to
both real and synthetic data are provided in \prettyref{sec:Simulations},
and \prettyref{sec:Conclusions} concludes the paper.

\vspace{-3mm}

\section{Problem Formulation\label{sec:Problem-Formulation}}

For simplicity of notations, we first introduce the AR(1) model. Suppose
a univariate time series $y_{1}$, $y_{2}$,$\ldots$, $y_{T}$ follows
an AR($1$) model 
\begin{equation}
y_{t}=\varphi_{0}+\varphi_{1}y_{t-1}+\varepsilon_{t},\label{eq:ar(1) model}
\end{equation}
where the innovations $\varepsilon_{t}$'s follow a zero-mean heavy-tailed
Student's $t$-distribution $\varepsilon_{t}\overset{i.i.d.}{\sim}t\left(0,\sigma^{2},\nu\right)$.
The Student's $t$-distribution is more heavy-tailed as the number
of degrees of freedom $\nu$ decreases. Note that the Gaussian distribution
is a special case of the Student's $t$-distribution with $\nu=+\text{\ensuremath{\infty}}.$

Given all the parameters $\varphi_{0}$, $\varphi_{1}$, $\sigma^{2}$
and $\nu$, the distribution of $y_{t}$ conditional on all the preceding
data $\mathcal{F}_{t-1}$, which consists of $y_{1}$,$y_{2}$, $\ldots$,$y_{t-1}$,
only depends on the previous sample $y_{t-1}$:

\begin{equation}
\begin{aligned} & p\left(y_{t}|\varphi_{0},\varphi_{1},\sigma^{2},\nu,\mathcal{F}_{t-1}\right)\\
 & =p\left(y_{t}|\varphi_{0},\varphi_{1},\sigma^{2},\nu,y_{t-1}\right)\\
 & =f_{t}\left(y_{t};\varphi_{0}+\varphi_{1}y_{t-1},\sigma^{2},\nu\right)\\
 & =\frac{\Gamma\left(\frac{\nu+1}{2}\right)}{\sqrt{\nu\pi}\sigma\Gamma\left(\frac{\nu}{2}\right)}\left(1+\frac{\left(y_{t}-\varphi_{0}-\varphi_{1}y_{t-1}\right)^{2}}{\nu\sigma^{2}}\right)^{-\frac{\nu+1}{2}},
\end{aligned}
\end{equation}
where $f_{t}\left(\cdot\right)$ denotes the probability density function
(pdf) of a Student's $t$-distribution.

In practice, a certain sample $y_{t}$ may be missing due to various
reasons, and it is denoted by $y_{t}=\mathsf{NA}$ (not available).
Here we assume that the missing-data mechanism is ignorable, i.e.,
the missing does not depend on the value \cite{little2002statistical}.
Suppose we have an observation of this time series with $D$ missing
blocks as follows: 
\[
\begin{aligned} & y_{1},\ldots,y_{t_{1}},\mathsf{NA},\ldots,\mathsf{NA},y_{t_{1}+n_{1}+1},\ldots y_{t_{d}},\mathsf{NA},\ldots,\mathsf{NA},\\
 & y_{t_{d}+n_{d}+1},\ldots,y_{t_{D}},\mathsf{NA},\ldots,\mathsf{NA},y_{t_{D}+n_{D}+1},\ldots,y_{T},
\end{aligned}
\]
where, in the $d$-th missing block, there are $n_{d}$ missing samples
$y_{t_{d}+1}$,$\ldots$,$y_{t_{d}+n_{d}}$, which are surrounded
from the left and the right by the two observed data $y_{t_{d}}$
and $y_{t_{d}+n_{d}+1}$. We set for convenience $t_{0}=0$ and $n_{0}=0$.
Let us denote the set of the indexes of the observed values by $C_{\mathsf{o}}$,
and the set of the indexes of the missing values by $C_{\mathsf{m}}$.
Also denote $\mathbf{y}=\left(y_{t},\thinspace1\leq t\leq T\right)$,
$\mathbf{y}_{\mathsf{o}}=\left(y_{t},\thinspace t\in C_{\mathsf{\mathsf{o}}}\right)$,
and $\mathbf{y}_{\mathsf{m}}=\left(y_{t},\thinspace t\in C_{\mathsf{m}}\right)$
.

Let us assume $\boldsymbol{\theta}=\left(\varphi_{0},\varphi_{1},\sigma^{2},\nu\right)\in\Theta$
with $\Theta=\left\{ \boldsymbol{\theta}|\sigma^{2}>0,\thinspace\nu>0\right\} .$
Ignoring the marginal distribution of $y_{1}$, the log-likelihood
of the observed data is 
\begin{align}
l\left(\boldsymbol{\theta};\mathbf{y}_{\mathsf{o}}\right)= & \log\left(\int p\left(\mathbf{y};\boldsymbol{\theta}\right)\mathsf{d}\mathbf{y}_{\mathsf{m}}\right)\nonumber \\
= & \log\left(\int\prod_{t=2}^{T}p\left(y_{t}|\boldsymbol{\theta},\mathcal{F}_{t-1}\right)\mathsf{d}\mathbf{y}_{\mathsf{m}}\right)\label{eq:objective funtion}\\
= & \log\left(\int\prod_{t=2}^{T}f_{t}\left(y_{t};\varphi_{0}+\varphi_{1}y_{t-1},\sigma^{2},\nu\right)\mathsf{d}\mathbf{y}_{\mathsf{m}}\right).\nonumber 
\end{align}
Then the maximum likelihood (ML) estimation problem for $\boldsymbol{\theta}$
can be formulated as 
\begin{equation}
\begin{aligned}\mathsf{\underset{\boldsymbol{\theta}\in\Theta}{maximize}} & \thinspace\thinspace\thinspace l\left(\boldsymbol{\theta};\mathbf{y}_{\mathsf{o}}\right).\end{aligned}
\label{eq:problem formulation}
\end{equation}

The integral in \eqref{eq:objective funtion} has no closed-form expression,
thus, the objective function is very complicated, and we cannot solve
the optimization problem directly. In order to deal with this, we
resort to the EM framework, which circumvents such difficulty by optimizing
a sequence of simpler approximations of the original objective function
instead.

\vspace{-3mm}

\section{EM and Its Stochastic Variants\label{sec:EM-and-Its-Stochastic-Variants}}

The EM algorithm is a general iterative algorithm to solve ML estimation
problems with missing data or latent data. More specifically, given
the observed data $\mathbf{X}$ generated from a statistical model
with unknown parameter $\boldsymbol{\theta}$, the ML estimator of
the parameter $\boldsymbol{\theta}$ is defined as the maximizer of
the likelihood of the observed data 
\begin{equation}
l\left(\mathbf{X};\boldsymbol{\theta}\right)=\log p(\mathbf{X}|\boldsymbol{\theta}).\label{eq:log-likelihood of EM}
\end{equation}
In practice, it often occurs that $l\left(\mathbf{X};\boldsymbol{\theta}\right)$
does not have manageable expression due to the missing data or latent
data $\mathbf{Z}$, while the likelihood of complete data $p(\mathbf{X},\mathbf{Z}|\boldsymbol{\theta})$
has a manageable expression. This is when the EM algorithm can help.
The EM algorithm seeks to find the ML estimates by iteratively applying
these two steps \cite{dempster1977maximum}: 
\begin{description}
\item [{(E)}] Expectation: calculate the expected log-likelihood of the
complete data set $\left(\mathbf{X},\mathbf{Z}\right)$ with respect
to the current conditional distribution of $\mathbf{Z}$ given $\mathbf{X}$
and the current estimate of the parameter $\mathbf{\boldsymbol{\theta}}^{\left(k\right)}$:
\begin{equation}
Q\left(\boldsymbol{\theta}|\boldsymbol{\theta}^{\left(k\right)}\right)=\int\log p\left(\mathbf{X},\thinspace\mathbf{Z}|\boldsymbol{\theta}\right)p\left(\mathbf{Z}|\mathbf{X},\boldsymbol{\theta}^{\left(k\right)}\right)\mathsf{d}\mathbf{Z},\label{eq:q function}
\end{equation}
where $k$ is the iteration number. 
\item [{(M)}] Maximization: find the new estimate 
\begin{equation}
\mathbf{\boldsymbol{\theta}}^{\left(k+1\right)}=\underset{\boldsymbol{\theta}}{\arg\max\thinspace}Q\left(\boldsymbol{\theta}|\mathbf{\boldsymbol{\theta}}^{\left(k\right)}\right).\label{eq:m-step}
\end{equation}
\end{description}
The sequence $\left\{ l\left(\mathbf{X};\boldsymbol{\theta}^{\left(k\right)}\right)\right\} $
generated by the EM algorithm is non-decreasing, and the limit points
of the sequence $\left\{ \mathbf{\boldsymbol{\theta}}^{\left(k\right)}\right\} $
are proven to be the stationary points of the observed data log-likelihood
under mild regularity conditions \cite{wu1983convergence}. In fact,
the EM algorithm is a particular choice of the more general majorization-minimization
algorithm \cite{sun2016majorization}.

However, in some applications of the EM algorithm, the expectation
in the E step cannot be obtained in closed-form. To deal with this,
Wei and Tanner proposed the Monte Carlo EM (MCEM) algorithm, in which
the expectation is computed by a Monte Carlo approximation based on
a large number of independent simulations of the missing data \cite{wei1990monte}.
The MCEM algorithm is computationally very intensive.

In order to reduce the amount of simulations required by the MCEM
algorithm, the stochastic approximation EM (SAEM) algorithm replaces
the E step of the EM algorithm by a stochastic approximation procedure,
which approximates the expectation by combining new simulations with
the previous ones \cite{delyon1999convergence}. At iteration $k$,
the SAEM proceeds as follows: 
\begin{description}
\item [{(E-S1)}] Simulation: generate $L$ realizations $\mathbf{Z}^{\left(k,l\right)}$
$\left(l=1,2\ldots,L\right)$ from the conditional distribution $p\left(\mathbf{Z}|\mathbf{X},\boldsymbol{\theta}^{\left(k\right)}\right)$ 
\item [{(E-A)}] Stochastic approximation: update $\hat{Q}\left(\boldsymbol{\theta}|\boldsymbol{\theta}^{\left(k\right)}\right)$
according to 
\begin{equation}
\begin{aligned} & \hat{Q}\left(\boldsymbol{\theta}|\boldsymbol{\theta}^{\left(k\right)}\right)\\
 & =\hat{Q}\left(\boldsymbol{\theta}|\boldsymbol{\theta}^{\left(k-1\right)}\right)+\gamma^{\left(k\right)}\biggl(\frac{1}{L}\sum_{l=1}^{L}\log p\left(\mathbf{X},\thinspace\mathbf{Z}^{\left(k,l\right)}|\boldsymbol{\theta}\right)\\
 & \hspace{10.5em}-\hat{Q}\left(\boldsymbol{\theta}|\boldsymbol{\theta}^{\left(k-1\right)}\right)\biggr),
\end{aligned}
\end{equation}
where $\left\{ \gamma^{\left(k\right)}\right\} $ is a decreasing
sequence of positive step sizes. 
\item [{(M)}] Maximization: find the new estimate 
\end{description}
\begin{equation}
\mathbf{\boldsymbol{\theta}}^{\left(k+1\right)}=\underset{\boldsymbol{\theta}}{\arg\max\thinspace}\hat{Q}\left(\boldsymbol{\theta}|\boldsymbol{\theta}^{\left(k\right)}\right).\label{eq:m-step-1}
\end{equation}
The SAEM requires a smaller amount of samples per iteration due to
the recycling of the previous simulations. A small value of $L$ is
enough to ensure satisfying results \cite{kuhn2005maximum}.

When the conditional distribution is very complicated, and the simulation
step (E-S1) of the SAEM cannot be directly performed, Kuhn and Lavielle
proposed to combine the SAEM algorithm with a Markov Chain Monte Carlo
(MCMC) procedure, which yields the SAEM-MCMC algorithm \cite{kuhn2004coupling}.
Assume the conditional distribution $p\left(\mathbf{Z}|\mathbf{X},\boldsymbol{\theta}\right)$
is the unique stationary distribution of the transition probability
density function $\Pi_{\boldsymbol{\theta}}$, the simulation step
of the SAEM is replaced with 
\begin{description}
\item [{(E-S2)}] Simulation: draw realizations $\mathbf{Z}^{\left(k,l\right)}$
$\left(l=1,2\ldots,L\right)$ based on the transition probability
density function $\Pi_{\boldsymbol{\theta}^{\left(k\right)}}\left(\mathbf{Z}^{\left(k-1,l\right)},\cdot\right)$. 
\end{description}
For each $l$, the sequence $\left\{ \mathbf{Z}^{\left(k,l\right)}\right\} _{k\geq0}$
is a Markov chain with the transition probability density function
$\left\{ \Pi_{\boldsymbol{\theta}^{\left(k\right)}}\right\} .$ The
Markov Chain generation mechanism needs to be well designed so that
the sampling is efficient and the computational cost is not too high.

\vspace{-1mm}

\section{SAEM-MCMC for Student's $t$ AR Model\label{sec:SAEM-MCMC-for-Student's}}

For the ML problem \eqref{eq:problem formulation}, if we only regard
$\mathbf{y}_{\mathsf{m}}$ as missing data and apply the EM type algorithm,
the resulting conditional distribution of the missing data is still
complicated, and it is difficult to maximize the expectation or the
approximated expectation of the complete data log-likelihood. Interestingly,
the Student's $t$-distribution can be regarded as a Gaussian mixture
\cite{liu1997ml}. Since $\varepsilon_{t}\sim t\left(0,\sigma^{2},\nu\right)$,
we can present it as a Gaussian mixture 
\begin{equation}
\varepsilon_{t}|\sigma^{2},\tau_{t}\thicksim\mathcal{N}\left(0,\frac{\sigma^{2}}{\tau_{t}}\right),\label{eq:gaussian mixture presentation of t-1-1}
\end{equation}
\begin{equation}
\tau_{t}\thicksim Gamma\left(\nu/2,\thinspace\nu/2\right),\label{eq:gaussian mixture presentation of t-2-1}
\end{equation}
where $\tau_{t}$ is the mixture weight. Denote $\boldsymbol{\tau}=\left\{ \tau_{t},\thinspace1<t\leq T\right\} $.
We can use the EM type algorithm to solve the above optimization problem
by regarding both $\mathbf{y}_{\mathsf{m}}$ and $\boldsymbol{\tau}$
as latent data, and $\mathbf{y}_{\mathsf{o}}$ as observed data.

The resulting complete data likelihood is 
\begin{equation}
\begin{aligned} & L\left(\boldsymbol{\theta};\mathbf{y},\boldsymbol{\tau}\right)\\
 & =p\left(\mathbf{y},\boldsymbol{\tau};\boldsymbol{\theta}\right)\\
 & =\prod_{t=2}^{T}\left\{ f_{N}\left(y_{t};\varphi_{0}+\varphi_{1}y_{t-1},\frac{\sigma^{2}}{\tau_{t}}\right)f_{g}\left(\tau_{t};\frac{\nu}{2},\frac{\nu}{2}\right)\right\} \\
 & =\prod_{t=2}^{T}\Biggl\{\frac{1}{\sqrt{2\pi\sigma^{2}/\tau_{t}}}\exp\left(-\frac{1}{2\sigma^{2}/\tau_{t}}\left(y_{t}-\varphi_{0}-\varphi_{1}y_{t-1}\right)^{2}\right)\\
 & \hspace{3em}\frac{\left(\frac{\nu}{2}\right)^{\frac{\nu}{2}}}{\Gamma\left(\frac{\nu}{2}\right)}\tau_{t}^{\frac{\nu}{2}-1}\exp\left(-\frac{\nu}{2}\tau_{t}\right)\Biggr\}\\
 & =\prod_{t=2}^{T}\frac{\left(\frac{\nu}{2}\right)^{\frac{\nu}{2}}\tau_{t}^{\frac{\nu-1}{2}}}{\Gamma\left(\frac{\nu}{2}\right)\sqrt{2\pi\sigma^{2}}}\exp\left(-\frac{\tau_{t}}{2\sigma^{2}}\left(y_{t}-\varphi_{0}-\varphi_{1}y_{t-1}\right)^{2}-\frac{\nu}{2}\tau_{t}\right),
\end{aligned}
\label{eq:complete data likelihood for ar(1)}
\end{equation}
where $f_{N}\left(\cdot\right)$ and $f_{g}\left(\cdot\right)$ denote
the pdf's of the Normal (Gaussian) and gamma distributions, respectively.
Through some simple derivation, it is observed that the likelihood
of complete data belongs to the curved exponential family \cite{DasGupta2011},
i.e., the pdf can be written as 
\begin{equation}
L\left(\boldsymbol{\theta};\mathbf{y},\boldsymbol{\tau}\right)=h\left(\mathbf{y},\boldsymbol{\tau}\right)\exp\left(-\psi\left(\boldsymbol{\theta}\right)+\left\langle \mathbf{s}\left(\mathbf{y}_{\mathsf{o}},\mathbf{y}_{\mathsf{m}},\boldsymbol{\tau}\right),\boldsymbol{\phi}\left(\boldsymbol{\theta}\right)\right\rangle \right),\label{eq:complete data likelihood}
\end{equation}
where $\left\langle \cdot,\cdot\right\rangle $ is the inner product,
\vspace{-2mm}
 
\begin{equation}
h\left(\mathbf{y},\boldsymbol{\tau}\right)=\prod_{t=2}^{T}\tau_{t}^{-\frac{1}{2}},
\end{equation}
\begin{equation}
\begin{aligned}\psi\left(\boldsymbol{\theta}\right)= & -\left(T-1\right)\Biggl\{\frac{\nu}{2}\log\left(\frac{\nu}{2}\right)-\log\left(\Gamma\left(\frac{\nu}{2}\right)\right)\\
 & \hspace{5em}-\frac{1}{2}\log\left(\sigma^{2}\right)-\frac{1}{2}\log\left(2\pi\right)\Biggr\},
\end{aligned}
\label{eq:psi}
\end{equation}
\begin{equation}
\boldsymbol{\phi}\left(\boldsymbol{\theta}\right)=\left[\frac{\nu}{2},\thinspace-\frac{1}{2\sigma^{2}},\thinspace-\frac{\varphi_{0}^{2}}{2\sigma^{2}},\thinspace-\frac{\varphi_{1}^{2}}{2\sigma^{2}},\thinspace\frac{\varphi_{0}}{\sigma^{2}},\thinspace\frac{\varphi_{1}}{\sigma^{2}},\thinspace-\frac{\varphi_{0}\varphi_{1}}{\sigma^{2}}\right],\label{eq:phi}
\end{equation}
and the minimal sufficient statistics 
\begin{align}
\mathbf{s}\left(\mathbf{y}_{\mathsf{o}},\mathbf{y}_{\mathsf{m}},\boldsymbol{\tau}\right)= & \Biggl[\sum_{t=2}^{T}\left(\log\left(\tau_{t}\right)-\tau_{t}\right),\sum_{t=2}^{T}\tau_{t}y_{t}^{2},\sum_{t=2}^{T}\tau_{t},\sum_{t=2}^{T}\tau_{t}y_{t-1}^{2},\nonumber \\
 & \sum_{t=2}^{T}\tau_{t}y_{t},\sum_{t=2}^{T}\tau_{t}y_{t}y_{t-1},\sum_{t=2}^{T}\tau_{t}y_{t-1}\Biggr].\label{eq:minimal sufficient statistics}
\end{align}

Then the expectation of the complete data log-likelihood can be expressed
as

\vspace{-4mm}
 
\begin{align}
 & Q\left(\boldsymbol{\theta}|\boldsymbol{\theta}^{\left(k\right)}\right)\nonumber \\
 & =\iint\log\left(L\left(\boldsymbol{\theta};\mathbf{y},\boldsymbol{\tau}\right)\right)p\left(\mathbf{y}_{\mathsf{m}},\boldsymbol{\tau}|\mathbf{y}_{\mathsf{o}};\boldsymbol{\theta}^{\left(k\right)}\right)\mathsf{d}\mathbf{y}_{\mathsf{m}}\mathsf{d}\boldsymbol{\tau}\nonumber \\
 & =\iint\log\biggl(h\left(\mathbf{y},\boldsymbol{\tau}\right)\exp\Bigl(-\psi\left(\boldsymbol{\theta}\right)+\Bigl\langle\mathbf{s}\left(\mathbf{y}_{\mathsf{o}},\mathbf{y}_{\mathsf{m}},\boldsymbol{\tau}\right),\boldsymbol{\phi}\left(\boldsymbol{\theta}\right)\Bigr\rangle\Bigr)\biggr)\nonumber \\
 & \hspace{3em}\times p\left(\mathbf{y}_{\mathsf{m}},\boldsymbol{\tau}|\mathbf{y}_{\mathsf{o}};\boldsymbol{\theta}^{\left(k\right)}\right)\mathsf{d}\mathbf{y}_{\mathsf{m}}\mathsf{d}\boldsymbol{\tau}\nonumber \\
 & =\iint\log\left(h\left(\mathbf{y},\boldsymbol{\tau}\right)\right)p\left(\mathbf{y}_{\mathsf{m}},\boldsymbol{\tau}|\mathbf{y}_{\mathsf{o}};\boldsymbol{\theta}^{\left(k\right)}\right)\mathsf{d}\mathbf{y}_{\mathsf{m}}\mathsf{d}\boldsymbol{\tau}\nonumber \\
 & \hspace{1.2em}-\psi\left(\boldsymbol{\theta}\right)+\Bigl\langle\iint\mathbf{s}\left(\mathbf{y}_{\mathsf{o}},\mathbf{y}_{\mathsf{m}},\boldsymbol{\tau}\right)p\left(\mathbf{y}_{\mathsf{m}},\boldsymbol{\tau}|\mathbf{y}_{\mathsf{o}};\boldsymbol{\theta}^{\left(k\right)}\right)\mathsf{d}\mathbf{y}_{\mathsf{m}}\mathsf{d}\boldsymbol{\tau},\nonumber \\
 & \hspace{1.2em}\boldsymbol{\phi}\left(\boldsymbol{\theta}\right)\Bigr\rangle\nonumber \\
 & =-\psi\left(\boldsymbol{\theta}\right)+\left\langle \bar{\mathbf{s}}\left(\boldsymbol{\theta}^{\left(k\right)}\right),\boldsymbol{\phi}\left(\boldsymbol{\theta}\right)\right\rangle +const.,\label{eq:expectation of complete data log-likelihood}
\end{align}
where 
\begin{equation}
\bar{\mathbf{s}}\left(\boldsymbol{\theta}^{\left(k\right)}\right)=\iint\mathbf{s}\left(\mathbf{y}_{\mathsf{o}},\mathbf{y}_{\mathsf{m}},\boldsymbol{\tau}\right)p\left(\mathbf{y}_{\mathsf{m}},\boldsymbol{\tau}|\mathbf{y}_{\mathsf{o}};\boldsymbol{\theta}^{\left(k\right)}\right)\mathsf{d}\mathbf{y}_{\mathsf{m}}\mathsf{d}\boldsymbol{\tau}.
\end{equation}
The EM algorithm is conveniently simplified by utilizing the properties
of the exponential family. The E step of the EM algorithm is reduced
to the calculation of the expected minimal sufficient statistics $\bar{\mathbf{s}}\left(\boldsymbol{\theta}^{\left(k\right)}\right)$,
and the M step is reduced to the maximization of the function \eqref{eq:expectation of complete data log-likelihood}.

\subsection{E step}

The conditional distribution of $\mathbf{y}_{\mathsf{m}}$ and $\boldsymbol{\tau}$
given $\mathbf{y}_{\mathsf{o}}$ and $\boldsymbol{\theta}$ is:

\begin{align}
 & p\left(\mathbf{y}_{\mathsf{m}},\boldsymbol{\tau}|\mathbf{y}_{\mathsf{o}};\boldsymbol{\theta}\right)\nonumber \\
 & =\frac{p\left(\mathbf{y},\boldsymbol{\tau};\boldsymbol{\theta}\right)}{p\left(\mathbf{y}_{\mathsf{o}};\boldsymbol{\theta}\right)}\nonumber \\
 & =\frac{p\left(\mathbf{y},\boldsymbol{\tau};\boldsymbol{\theta}\right)}{\iint p\left(\mathbf{y},\boldsymbol{\tau};\boldsymbol{\theta}\right)\mathsf{d}\mathbf{y}_{\mathsf{m}}\mathsf{d\boldsymbol{\tau}}}\nonumber \\
 & \propto p\left(\mathbf{y},\boldsymbol{\tau};\boldsymbol{\theta}\right)\nonumber \\
 & =\prod_{t=2}^{T}\frac{\left(\frac{\nu}{2}\right)^{\frac{\nu}{2}}\tau_{t}^{\frac{\nu-1}{2}}}{\Gamma\left(\frac{\nu}{2}\right)\sqrt{2\pi\sigma^{2}}}\exp\left(-\frac{\tau_{t}}{2\sigma^{2}}\left(y_{t}-\varphi_{0}-\varphi_{1}y_{t-1}\right)^{2}-\frac{\nu}{2}\tau_{t}\right)\nonumber \\
 & \propto\prod_{t=2}^{T}\tau_{t}^{\frac{\nu-1}{2}}\exp\biggl(-\frac{\tau_{t}}{2\sigma^{2}}\left(y_{t}-\varphi_{0}-\varphi_{1}y_{t-1}\right)^{2}-\frac{\nu}{2}\tau_{t}\biggr)\text{.}\label{eq:proportional term}
\end{align}
Since the integral $\iint p\left(\mathbf{y},\boldsymbol{\tau};\boldsymbol{\theta}\right)\mathsf{d}\mathbf{y}_{\mathsf{m}}\mathsf{d\boldsymbol{\tau}}$
does not have a closed-from expression, we only know $p\left(\mathbf{y}_{\mathsf{m}},\boldsymbol{\tau}|\mathbf{y}_{\mathsf{o}};\boldsymbol{\theta}\right)$
up to a scalar. In addition, the proportional term is complicated,
and we cannot get closed-form expression for the conditional expectations
$\bar{\mathbf{s}}\bigl(\boldsymbol{\theta}^{\left(k\right)}\bigr)$
or $Q\bigl(\boldsymbol{\theta}|\boldsymbol{\theta}^{\left(k\right)}\bigr)$.
Therefore, we resort to the SAEM-MCMC algorithm, which generates samples
from the conditional distribution using a Markov chain process, and
approximates the expectation $\bar{\mathbf{s}}\bigl(\boldsymbol{\theta}^{\left(k\right)}\bigr)$
and $Q\bigl(\boldsymbol{\theta}|\boldsymbol{\theta}^{\left(k\right)}\bigr)$
by a stochastic approximation.

We propose to use the Gibbs sampling method to generate the Markov
chains. The Gibbs sampler divides the latent variables $\left(\mathbf{y}_{\mathsf{m}},\boldsymbol{\tau}\right)$
into two blocks $\boldsymbol{\tau}$ and $\mathbf{y}_{\mathsf{m}}$,
and then generates a Markov chain of samples from the distribution
$p\left(\mathbf{y}_{\mathsf{m}},\boldsymbol{\tau}|\mathbf{y}_{\mathsf{o}};\boldsymbol{\theta}\right)$
by drawing realizations from its conditional distributions $p\left(\boldsymbol{\tau}|\mathbf{y}_{\mathsf{m}},\mathbf{y}_{\mathsf{o}};\boldsymbol{\theta}\right)$
and $p\left(\mathbf{y}_{\mathsf{m}}|\boldsymbol{\tau},\mathbf{y}_{\mathsf{o}};\boldsymbol{\theta}\right)$
alternatively. More specifically, at iteration $k$, given the current
estimate $\boldsymbol{\theta}^{\left(k\right)},$ the Gibbs sampler
starts with $\left(\boldsymbol{\tau}^{\left(k-1,l\right)},\mathbf{y}_{\mathsf{m}}^{\left(k-1,l\right)}\right)$
$\left(l=1,2\ldots,L\right)$ and generate the next sample $\left(\boldsymbol{\tau}^{\left(k,l\right)},\mathbf{y}_{\mathsf{m}}^{\left(k,l\right)}\right)$
via the following scheme: 
\begin{itemize}
\item sample $\boldsymbol{\tau}^{\left(k,l\right)}$ from $p\left(\boldsymbol{\tau}|\mathbf{y}_{\mathsf{m}}^{\left(k-1,l\right)},\mathbf{y}_{\mathsf{o}};\boldsymbol{\theta}^{\left(k\right)}\right)$, 
\item sample $\mathbf{y}_{\mathsf{m}}^{\left(k,l\right)}$ from $p\left(\mathbf{y}_{\mathsf{m}}|\boldsymbol{\tau}^{\left(k,l\right)},\mathbf{y}_{\mathsf{o}};\boldsymbol{\theta}^{\left(k\right)}\right)$. 
\end{itemize}
Then the expected minimal sufficient statistics $\bar{\mathbf{s}}\bigl(\boldsymbol{\theta}^{\left(k\right)}\bigr)$
and the expected complete data likelihood $Q\bigl(\boldsymbol{\theta}|\boldsymbol{\theta}^{\left(k\right)}\bigr)$
are approximated by 
\begin{equation}
\hat{\mathbf{s}}^{\left(k\right)}=\hat{\mathbf{s}}^{\left(k-1\right)}+\gamma^{\left(k\right)}\left(\frac{1}{L}\sum_{l=1}^{L}\mathbf{s}\left(\mathbf{y}_{\mathsf{o}},\mathbf{y}_{\mathsf{m}}^{\left(k,l\right)},\boldsymbol{\tau}^{\left(k,l\right)}\right)-\hat{\mathbf{s}}^{\left(k-1\right)}\right),\label{eq:s approximation}
\end{equation}
\begin{equation}
\hat{Q}\left(\boldsymbol{\theta},\hat{\mathbf{s}}^{\left(k\right)}\right)=-\psi\left(\boldsymbol{\theta}\right)+\left\langle \hat{\mathbf{s}}^{\left(k\right)},\boldsymbol{\phi}\left(\boldsymbol{\theta}\right)\right\rangle +const.\label{eq:q approximation}
\end{equation}

Lemmas \ref{lem:lemma1} and \ref{lem:lemma2} give the two conditional
distributions $p\left(\boldsymbol{\tau}|\mathbf{y}_{\mathsf{m}},\mathbf{y}_{\mathsf{o}};\boldsymbol{\theta}\right)$
and $p\left(\mathbf{y}_{\mathsf{m}}|\boldsymbol{\tau},\mathbf{y}_{\mathsf{o}};\boldsymbol{\theta}\right)$.
Basically, to sample from them, we just need to draw realizations
from certain Gaussian distributions and gamma distributions, which
is simple. Based on the above sampling scheme, we can get the transition
probability density function of the Markov chain as follows: 
\begin{equation}
\Pi_{\boldsymbol{\theta}}\left(\mathbf{y}_{\mathsf{m}},\boldsymbol{\tau},\thinspace\mathbf{y}_{\mathsf{m}}',\boldsymbol{\tau}'\right)=p\left(\boldsymbol{\tau}'|\mathbf{y}_{\mathsf{m}},\mathbf{y}_{\mathsf{o}};\boldsymbol{\theta}\right)p\left(\mathbf{y}_{\mathsf{m}}'|\boldsymbol{\tau}',\mathbf{y}_{\mathsf{o}};\boldsymbol{\theta}\right).\label{eq:transition probability}
\end{equation}

\begin{lem}
\textup{\emph{\label{lem:lemma1}Given $\mathbf{y}_{\mathsf{m}}$,
$\mathbf{y}_{\mathsf{o}}$, and}}\textup{ }\textup{\emph{$\boldsymbol{\theta}$,
the mixture weights $\left\{ \tau_{t}\right\} $ are independent from
each other}}\textup{, }\textup{\emph{i.e., }}\textup{
\begin{equation}
p\left(\boldsymbol{\tau}|\mathbf{y}_{\mathsf{m}},\mathbf{y}_{\mathsf{o}};\boldsymbol{\theta}\right)=\prod_{t=2}^{T}p\left(\tau_{t}|\mathbf{y}_{\mathsf{m}},\mathbf{y}_{\mathsf{o}};\boldsymbol{\theta}\right).\label{eq:conditional_distribution_of_tau}
\end{equation}
}\textup{\emph{In addition, }}\textup{$\tau_{t}$ }\textup{\emph{follows
a gamma distribution:}}\textup{ 
\begin{equation}
\begin{aligned} & \tau_{t}|\mathbf{y}_{\mathsf{m}},\mathbf{y}_{\mathsf{o}};\boldsymbol{\theta}\\
 & \sim\ Gamma\left(\frac{\nu+1}{2},\thinspace\frac{\left(y_{t}-\varphi_{0}-\varphi_{1}y_{t-1}\right)^{2}/\sigma^{2}+\nu}{2}\right).
\end{aligned}
\end{equation}
} 
\end{lem}
\begin{IEEEproof}
See Appendix \ref{subsec:Proof-for-Lemma 1}. 
\end{IEEEproof}
\begin{lem}
\textup{\emph{\label{lem:lemma2}Given $\boldsymbol{\tau}$, $\mathbf{y}_{\mathsf{o}}$,
and $\boldsymbol{\theta}$, the missing blocks }}\textup{$\mathbf{y}_{d}=\left[y_{t_{d}+1},y_{t_{d}+2},\ldots,y_{t_{d}+n_{d}}\right]^{T}$}\textup{\emph{,}}\textup{
}\textup{\emph{where $d=1,2,\ldots,$$D$, are independent from each
other, i.e., 
\begin{equation}
p\left(\mathbf{y}_{\mathsf{m}}|\boldsymbol{\tau},\mathbf{y}_{\mathsf{o}};\boldsymbol{\theta}\right)=\prod_{d=1}^{D}p\left(\mathbf{y}_{d}|\boldsymbol{\tau},\mathbf{y}_{\mathsf{o}};\boldsymbol{\theta}\right).\label{eq:conditional_distribuation_of_y_m}
\end{equation}
In addition, the conditional distribution of $\mathbf{y}_{d}$ only
depends on the two nearest observed samples $y_{t_{d}}$ and $y_{t_{d}+n_{d}+1}$
with 
\begin{equation}
\mathbf{y}_{d}|\boldsymbol{\tau},\mathbf{y}_{\mathsf{o}};\boldsymbol{\theta}\sim\mathit{\mathcal{N}}\left(\mathbf{\boldsymbol{\mu}}_{d},\boldsymbol{\Sigma}_{d}\right),\label{eq:distribution of missing block}
\end{equation}
}}where the $i$-th component of $\mathbf{\boldsymbol{\mu}}_{d}$
\begin{equation}
\begin{aligned}\mathbf{\mu}_{d\left(i\right)}= & \sum_{q=0}^{i-1}\varphi_{1}^{q}\varphi_{0}+\varphi_{1}^{i}y_{t_{d}}+\frac{\sum_{q=1}^{i}\frac{\varphi_{1}^{i-2q}}{\tau_{t_{d}+q}}}{\sum_{q=1}^{n_{d}+1}\frac{\varphi_{1}^{n_{d}+1-2q}}{\tau_{t_{d}+q}}}\\
 & \times\left(y_{t_{d}+n_{d}+1}-\sum_{q=0}^{n_{d}}\varphi_{1}^{q}\varphi_{0}-\varphi_{1}^{n_{d}+1}y_{t_{d}}\right),
\end{aligned}
\label{eq:mean of missing block}
\end{equation}
and the component in the $i$-th column and the $j$-th row of $\boldsymbol{\Sigma}_{d}$\emph{
}\textup{ 
\begin{equation}
\begin{aligned} & \Sigma_{d\left(i,j\right)}\\
 & =\left(\sum_{q=1}^{\min\left(i,j\right)}\frac{\varphi_{1}^{i+j-2q}}{\tau_{t_{d}+q}}-\frac{\left(\sum_{q=1}^{i}\frac{\varphi_{1}^{i-2q}}{\tau_{t_{d}+q}}\right)\left(\sum_{q=1}^{j}\frac{\varphi_{1}^{j-2q}}{\tau_{t_{d}+q}}\right)}{\sum_{q=1}^{n_{d}+1}\frac{\varphi_{1}^{-2q}}{\tau_{t_{d}+q}}}\right)\sigma^{2},
\end{aligned}
\label{eq:covariance of missing block}
\end{equation}
}where the sums of geometric progressions in \textup{$\mathbf{\mu}_{d\left(i\right)}$}
can be simplified as 
\begin{equation}
\sum_{q=0}^{i-1}\varphi_{1}^{q}\varphi_{0}=\begin{cases}
i\varphi_{0}, & \varphi_{1}=1,\\
\frac{\varphi_{0}\left(\varphi_{1}^{i}-1\right)}{\varphi_{1}-1}, & \varphi_{1}\neq1,
\end{cases}
\end{equation}
and\textup{ 
\begin{equation}
\sum_{q=0}^{n_{d}}\varphi_{1}^{q}\varphi_{0}=\begin{cases}
\left(n_{d}+1\right)\varphi_{0}, & \varphi_{1}=1,\\
\frac{\varphi_{0}\left(\varphi_{1}^{n_{d}+1}-1\right)}{\varphi_{1}-1}, & \varphi_{1}\neq1.
\end{cases}
\end{equation}
} 
\end{lem}
\begin{IEEEproof}
See Appendix \ref{subsec:Proof-for-Lemma 2}.  
\end{IEEEproof}

\subsection{M step}

After obtaining the approximation $\hat{Q}\left(\boldsymbol{\theta},\hat{\mathbf{s}}^{\left(k\right)}\right)$
in \eqref{eq:q approximation}, we need to maximize it to update the
estimates. The function $\hat{Q}\left(\boldsymbol{\theta},\hat{\mathbf{s}}^{\left(k\right)}\right)$
can be rewritten as 
\begin{equation}
\begin{aligned} & \hat{Q}\left(\boldsymbol{\theta},\hat{\mathbf{s}}^{\left(k\right)}\right)\\
 & =-\psi\left(\boldsymbol{\theta}\right)+\left\langle \hat{\mathbf{s}}^{\left(k\right)},\boldsymbol{\phi}\left(\boldsymbol{\theta}\right)\right\rangle +const.\\
 & =\left(T-1\right)\left\{ \frac{\nu}{2}\log\left(\frac{\nu}{2}\right)-\log\left(\Gamma\left(\frac{\nu}{2}\right)\right)-\frac{1}{2}\log\left(\sigma^{2}\right)\right\} \\
 & \hspace{1.5em}+\frac{\nu}{2}\hat{s}_{1}^{\left(k\right)}-\frac{\hat{s}_{2}^{\left(k\right)}}{2\sigma^{2}}-\frac{\varphi_{0}^{2}\hat{s}_{3}^{\left(k\right)}}{2\sigma^{2}}-\frac{\varphi_{1}^{2}\hat{s}_{4}^{\left(k\right)}}{2\sigma^{2}}+\frac{\varphi_{0}\hat{s}_{5}^{\left(k\right)}}{\sigma^{2}}+\frac{\varphi_{1}\hat{s}_{6}^{\left(k\right)}}{\sigma^{2}}\\
 & \hspace{1.5em}-\frac{\varphi_{0}\varphi_{1}\hat{s}_{7}^{\left(k\right)}}{\sigma^{2}}+const,
\end{aligned}
\label{eq:approximated_q}
\end{equation}
where $\hat{s}_{i}^{\left(k\right)}$ ($i=1,2,\ldots,7$) is the $i$-th
component of $\hat{\mathbf{s}}^{\left(k\right)}$.

The optimization of $\varphi_{0}$, $\varphi_{1}$, and $\sigma^{2}$
is decoupled from the optimization of $\nu$. Setting the derivatives
of $\hat{Q}\left(\boldsymbol{\theta},\hat{\mathbf{s}}^{\left(k\right)}\right)$
with respect to to $\varphi_{0}$, $\varphi_{1}$, and $\sigma^{2}$
to $0$ gives 
\begin{equation}
\varphi_{0}^{\left(k+1\right)}=\frac{\hat{s}_{5}^{\left(k\right)}-\varphi_{1}^{\left(k+1\right)}\hat{s}_{7}^{\left(k\right)}}{\hat{s}_{3}^{\left(k\right)}},\label{eq:maximizer phi0}
\end{equation}
\begin{equation}
\varphi_{1}^{\left(k+1\right)}=\frac{\hat{s}_{3}^{\left(k\right)}\hat{s}_{6}^{\left(k\right)}-\hat{s}_{5}^{\left(k\right)}\hat{s}_{7}^{\left(k\right)}}{\hat{s}_{3}^{\left(k\right)}\hat{s}_{4}^{\left(k\right)}-\left(\hat{s}_{7}^{\left(k\right)}\right)^{2}},\label{eq:maximizer phi 1}
\end{equation}
and 
\begin{equation}
\begin{aligned}\left(\sigma^{\left(k+1\right)}\right)^{2}= & \frac{1}{T-1}\biggl(\hat{s}_{2}^{\left(k\right)}+\left(\varphi_{0}^{\left(k+1\right)}\right)^{2}\hat{s}_{3}^{\left(k\right)}+\left(\varphi_{1}^{\left(k+1\right)}\right)^{2}\hat{s}_{4}^{\left(k\right)}\\
 & \hspace{3em}-2\varphi_{0}^{\left(k+1\right)}\hat{s}_{5}^{\left(k\right)}-2\varphi_{1}^{\left(k+1\right)}\hat{s}_{6}^{\left(k\right)}\\
 & \hspace{3em}+2\varphi_{0}^{\left(k+1\right)}\varphi_{1}^{\left(k+1\right)}\hat{s}_{7}^{\left(k\right)}\biggr).
\end{aligned}
\label{eq:maximzer sigma}
\end{equation}
The $\nu^{\left(k+1\right)}$ can be found by: 
\begin{equation}
\nu^{\left(k+1\right)}=\underset{\nu>0}{\arg\max}\ f\left(\nu,\hat{s}_{1}^{\left(k\right)}\right)\label{eq:maximizer nu}
\end{equation}
with $f\left(\nu,\hat{s}_{1}^{\left(k\right)}\right)=\left\{ \frac{\nu}{2}\log\left(\frac{\nu}{2}\right)-\log\left(\Gamma\left(\frac{\nu}{2}\right)\right)\right\} +\frac{\nu\hat{s}_{1}^{\left(k\right)}}{2\left(T-1\right)}.$
According to Proposition 1 in \cite{liu1995ml}, $\nu^{\left(k+1\right)}$
always exists and is unique. As suggested in \cite{liu1995ml}, the
maximizer $\nu^{\left(k+1\right)}$ can be obtained by one-dimensional
search, such as half interval method \cite{carnahan1969applied}.

The resulting SAEM-MCMC algorithm is summarized in Algorithm 1.

\begin{algorithm}[tbh]
\label{SAEM:algo-1} \caption{SAEM-MCMC Algorithm for Student's $t$ AR(1)}
\begin{algorithmic}[1] \STATE Initialize $\boldsymbol{\theta}^{\left(0\right)}\in\Theta$,
$\hat{\mathbf{s}}^{\left(0\right)}=0$, $k=0,$ and $\mathbf{y}_{\mathsf{m}}^{\left(0,l\right)}$
for $l=1,2\ldots,L.$. \FOR{$k=1,2,\ldots$} \STATE Simulation:\FOR{$l=1,2\ldots,L$}\STATE
sample $\boldsymbol{\tau}^{\left(k,l\right)}$ from $p\left(\boldsymbol{\tau}|\mathbf{y}_{\mathsf{m}}^{\left(k-1,l\right)},\mathbf{y}_{\mathsf{o}};\boldsymbol{\theta}^{\left(k\right)}\right)$
using \prettyref{lem:lemma1}, \STATE sample $\mathbf{y}_{\mathsf{m}}^{\left(k,l\right)}$
for $p\left(\mathbf{y}_{\mathsf{m}}|\boldsymbol{\tau}^{\left(k,l\right)},\mathbf{y}_{\mathsf{o}};\boldsymbol{\theta}^{\left(k\right)}\right)$
using \prettyref{lem:lemma2}. \ENDFOR \STATE Stochastic approximation:
evaluate $\hat{\mathbf{s}}^{\left(k\right)}$ and $\hat{Q}\left(\boldsymbol{\theta},\hat{\mathbf{s}}^{\left(k\right)}\right)$
as in \eqref{eq:s approximation} and \eqref{eq:q approximation}
respectively. 
%
\STATE Maximization: update $\boldsymbol{\theta}^{\left(k+1\right)}$
as in \eqref{eq:maximizer phi0}, \eqref{eq:maximizer phi 1}, \eqref{eq:maximzer sigma}
and \eqref{eq:maximizer nu}. \IF{stopping criteria is met}\STATE
terminate loop \ENDIF \ENDFOR \end{algorithmic} 
\end{algorithm}

\subsection{Particular Cases}

In cases where some parameters in $\boldsymbol{\theta}$ are known,
we just need to change the updates in M-step accordingly, and the
simulation and approximation steps remain the same. For example, if
we know that the time series is zero mean \cite{christmas2011robust,choong2009autoregressive},
i.e., $\varphi_{0}=0$, then the update for $\varphi_{0}^{\left(k+1\right)}$
and $\varphi_{1}^{\left(k+1\right)}$ should be replaced with 
\begin{equation}
\varphi_{0}^{\left(k+1\right)}=0,\label{eq:maximizer phi0-1}
\end{equation}
and 
\begin{equation}
\varphi_{1}^{\left(k+1\right)}=\frac{\hat{s}_{6}^{\left(k\right)}}{\hat{s}_{4}^{\left(k\right)}},\label{eq:maximizer phi 1-1}
\end{equation}

If the time series is known to follow the random walk model \cite{dicesare2006imputation},
which is a special case of AR(1) model with $\varphi_{1}=1$, then
the update for $\varphi_{0}^{\left(k+1\right)}$ and $\varphi_{1}^{\left(k+1\right)}$
should be replaced with 
\begin{equation}
\varphi_{0}^{\left(k+1\right)}=\frac{\hat{s}_{5}^{\left(k\right)}-\hat{s}_{7}^{\left(k\right)}}{\hat{s}_{3}^{\left(k\right)}},\label{eq:maximizer phi0-2}
\end{equation}
and 
\begin{equation}
\varphi_{1}^{\left(k+1\right)}=1.\label{eq:maximizer phi 1-2}
\end{equation}

\subsection{Generalization to AR($p$)}

The above ML estimation method can be immediately generalized to the
Student's $t$ AR($p$) model: 
\begin{equation}
y_{t}=\varphi_{0}+\sum_{i=1}^{p}\varphi_{i}y_{t-i}+\varepsilon_{t},\label{eq:ar(p) model}
\end{equation}
where $\varepsilon_{t}\overset{i.i.d.}{\sim}t\left(0,\sigma^{2},\nu\right)$.
Similarly, we can apply the SAEM-MCMC algorithm to obtain the estimates
by considering $\boldsymbol{\tau}$ and $\mathbf{y}_{\mathsf{m}}$
as latent data, and $\mathbf{y}_{\mathsf{o}}$ as observed data. At
each iteration, we draw some realizations of $\boldsymbol{\tau}$
and $\mathbf{y}_{\mathsf{m}}$ from the conditional distribution $p\left(\mathbf{y}_{\mathsf{m}},\boldsymbol{\tau}|\mathbf{y}_{\mathsf{o}};\boldsymbol{\theta}^{\left(k\right)}\right)$
to approximate the expectation function $Q\left(\boldsymbol{\theta}|\boldsymbol{\theta}^{\left(k\right)}\right)$,
and maximize the approximation $\hat{Q}\left(\boldsymbol{\theta}|\boldsymbol{\theta}^{\left(k\right)}\right)$
to update the estimates. The main difference is that the conditional
distribution of the AR($p$) will become more complicated than that
of the AR($1$), since each sample of the AR($p$) has more dependence
on the previous samples. To deal with this challenge, when applying
the Gibbs sampling, we can divide the the latent data $\left(\mathbf{y}_{\mathsf{m}},\boldsymbol{\tau}\right)$
into more blocks, $\boldsymbol{\tau}$ as a block and each $y_{i\in C_{m}}$
as a block, so that the distribution of each block of latent variables
conditional on other latent variables will be easy to obtain and sample
from. For limit of space, we do not go into details here, and we will
consider this in our future work.

\section{Convergence\label{sec:Convergence}}

In this section, we provide theoretical guarantee for the convergence
of the proposed algorithm. The convergence of the simple deterministic
EM algorithm has been addressed by many different authors, starting
from the seminal work in \cite{dempster1977maximum}, to a more general
consideration in \cite{wu1983convergence}. However, the convergence
analysis of stochastic variants of the EM algorithm, like the MCEM,
SAEM and SAEM-MCMC algorithms, is challenging due to the randomness
of sampling. See \cite{chan1995monte,gu1998stochastic,fort2003convergence,delyon1999convergence,kuhn2004coupling,neath2013convergence}
for a more general overview of these stochastic EM algorithms and
their convergence analysis. Of specific interest, the authors in \cite{delyon1999convergence}
introduced the SAEM algorithm, and established the almost sure convergence
to the stationary points of the observed data likelihood under mild
additional conditions. The authors in \cite{kuhn2004coupling} coupled
the SAEM framework with an MCMC procedure, and they have given the
convergence conditions for the SAEM-MCMC algorithm when the complete
data likelihood belongs to the curved exponential family. The given
set of conditions in our case is as follows. 
\begin{description}
\item [{(M1)}] For any $\boldsymbol{\theta}\in\Theta,$ 
\begin{equation}
\int\int\|\mathbf{s}\left(\mathbf{y}_{\mathsf{o}},\mathbf{y}_{\mathsf{m}},\boldsymbol{\tau}\right)\|p\left(\mathbf{y}_{\mathsf{m}},\boldsymbol{\tau}|\mathbf{y}_{\mathsf{o}};\boldsymbol{\theta}\right)\mathsf{d}\mathbf{y}_{\mathsf{m}}\mathsf{d\boldsymbol{\tau}}<\infty.
\end{equation}
\item [{(M2)}] $\mathbf{\psi}\left(\boldsymbol{\theta}\right)$ and $\boldsymbol{\phi}\left(\boldsymbol{\theta}\right)$
are twice continuously differentiable on $\Theta$. 
\item [{(M3)}] The function 
\begin{equation}
\bar{\mathbf{s}}\left(\boldsymbol{\theta}\right)=\int\int\mathbf{s}\left(\mathbf{y}_{\mathsf{o}},\mathbf{y}_{\mathsf{m}},\boldsymbol{\tau}\right)p\left(\mathbf{y}_{\mathsf{m}},\boldsymbol{\tau}|\mathbf{y}_{\mathsf{o}};\boldsymbol{\theta}\right)\mathsf{d}\mathbf{y}_{\mathsf{m}}\mathsf{d\boldsymbol{\tau}}
\end{equation}
is continuously differentiable on $\Theta$. 
\item [{(M4)}] The objective function 
\begin{equation}
l\left(\boldsymbol{\theta};\mathbf{y}_{\mathsf{o}}\right)=\log\left(\int\int p\left(\mathbf{y},\boldsymbol{\tau};\boldsymbol{\theta}\right)\mathsf{d}\mathbf{y}_{\mathsf{m}}\mathsf{d\boldsymbol{\tau}}\right)
\end{equation}
is continuously differentiable on $\Theta$, and 
\end{description}
\begin{equation}
\partial_{\boldsymbol{\theta}}\int\int p\left(\mathbf{y},\boldsymbol{\tau};\boldsymbol{\theta}\right)\mathsf{d}\mathbf{y}_{m}\mathsf{d\boldsymbol{\tau}}=\int\int\partial_{\boldsymbol{\theta}}p\left(\mathbf{y},\boldsymbol{\tau};\boldsymbol{\theta}\right)\mathsf{d}\mathbf{y}_{m}\mathsf{d\boldsymbol{\tau}}.\label{eq:change the order of derivative and integral}
\end{equation}

\begin{description}
\item [{(M5)}] For $Q\left(\boldsymbol{\theta},\bar{\mathbf{s}}\right)=-\psi\left(\boldsymbol{\theta}\right)+\left\langle \bar{\mathbf{s}},\boldsymbol{\phi}\left(\boldsymbol{\theta}\right)\right\rangle +const.$,
there exists a function $\tilde{\boldsymbol{\theta}}\left(\bar{\mathbf{s}}\right)$
such that $\forall\bar{\mathbf{s}}$ and $\forall\boldsymbol{\theta}\in\Theta,$
$Q\left(\tilde{\boldsymbol{\theta}}\left(\bar{\mathbf{s}}\right),\bar{\mathbf{s}}\right)\geq Q\left(\boldsymbol{\theta},\bar{\mathbf{s}}\right).$
In addition, the function $\tilde{\boldsymbol{\theta}}\left(\bar{\mathbf{s}}\right)$
is continuously differentiable. 
\item [{(SAEM1)}] \hspace{1.5em}For all $k$, $\gamma^{\left(k\right)}\in\left[0,1\right]$,
$\sum_{k=1}^{\infty}\gamma^{\left(k\right)}=\infty$ and there exists
$\frac{1}{2}<\lambda\leq1$ such that $\sum_{k=1}^{\infty}\left(\gamma^{\left(k\right)}\right)^{1+\lambda}<\infty$. 
\item [{(SAEM2)}] \hspace{1.5em}$l\left(\boldsymbol{\theta};\mathbf{y}_{\mathsf{o}}\right)$
is $d$ times differentiable on $\Theta$, where $d=7$ is the dimension
of $\mathbf{s}\left(\mathbf{y}_{\mathsf{o}},\mathbf{y}_{\mathsf{m}},\boldsymbol{\tau}\right)$,
and $\tilde{\boldsymbol{\theta}}\left(\mathbf{s}\right)$ is $d$
times differentiable. 
\item [{(SAEM3)}] ~ 
\begin{enumerate}
\item The chain takes its values in a compact set $\Omega$. 
\item The $\mathbf{s}\left(\mathbf{y}_{\mathsf{o}},\mathbf{y}_{\mathsf{m}},\boldsymbol{\tau}\right)$
is bounded on $\Omega$, and the sequence $\left\{ \hat{\mathbf{s}}^{\left(k\right)}\right\} $
takes its values in a compact subset. 
\item For any compact subset $V$ of $\Theta$, there exists a real constant
$L$ such that for any $\left(\boldsymbol{\theta},\boldsymbol{\theta}'\right)$
in $V^{2}$ 
\begin{equation}
\begin{aligned} & \underset{\left(\mathbf{y}_{\mathsf{m}},\boldsymbol{\tau},\mathbf{y}_{\mathsf{m}}',\boldsymbol{\tau}'\right)\in\Omega^{2}}{\sup}\Bigl|\Pi_{\boldsymbol{\theta}}\left(\mathbf{y}_{\mathsf{m}},\boldsymbol{\tau},\thinspace\mathbf{y}_{\mathsf{m}}',\boldsymbol{\tau}'\right)\\
 & \hspace{6em}-\Pi_{\boldsymbol{\theta}'}\left(\mathbf{y}_{\mathsf{m}},\boldsymbol{\tau},\thinspace\mathbf{y}_{\mathsf{m}}',\boldsymbol{\tau}'\right)\Bigl|\\
 & \leq L|\boldsymbol{\theta}-\boldsymbol{\theta}'|\text{.}
\end{aligned}
\end{equation}
\item The transition probability $\Pi_{\boldsymbol{\theta}}$ generates
a uniformly ergodic chain whose invariant probability is the conditional
distribution $p\left(\mathbf{y}_{\mathsf{m}},\boldsymbol{\tau}|\mathbf{y}_{\mathsf{o}};\boldsymbol{\theta}\right)$. 
\end{enumerate}
\end{description}
In summary, the conditions (M1)-(M5) are all about the model, and
are conditions for the convergence of the deterministic EM algorithm.
The conditions (M1) and (M3) require the boundedness and continuous
differentiability of the expectation of the sufficient statistics.
The conditions (M2) and (M4) guarantee the continuous differentiability
of the complete data log-likelihood $l\left(\boldsymbol{\theta};\mathbf{y},\boldsymbol{\tau}\right)$,
the expectation of the complete data likelihood $Q\left(\boldsymbol{\theta}|\boldsymbol{\theta}^{\left(k\right)}\right)$,
and the observed data log-likelihood $l\left(\boldsymbol{\theta};\mathbf{y}_{\mathsf{o}}\right)$.
The condition (M5) indicates the existence of a global maximizer for
$Q\left(\boldsymbol{\theta},\bar{\mathbf{s}}\right)$.

The conditions (SAEM1)-(SAEM3) are additional requirements for the
SAEM-MCMC convergence. The condition (SAEM1) is about the step sizes
$\left\{ \gamma^{\left(k\right)}\right\} .$ This condition can be
easily satisfied by choosing the step sizes properly. It is recommended
to set $\gamma^{\left(k\right)}=1$ for 1$\leq k\leq K$ and $\gamma^{\left(k\right)}=\frac{1}{k-K}$
for $k\geq K+1,$ where $K$ is a positive integer, since the initial
guess $\boldsymbol{\theta}^{\left(0\right)}$ may be far from the
ML estimates we are looking for, and choosing the first $K$ step
sizes equal to 1 allows the sequence $\left\{ \boldsymbol{\theta}^{\left(k\right)}\right\} $
to have a large variation and then converge to a neighborhood of the
maximum likelihood \cite{kuhn2005maximum}. The condition (SAEM2)
requires $d=7$ times differentiability of $l\left(\boldsymbol{\theta};\mathbf{y}_{\mathsf{o}}\right)$
and $\hat{\boldsymbol{\theta}}\left(\hat{\mathbf{s}}^{\left(k\right)}\right)$.
The condition (SAEM3) imposes some constraints on the generated Markov
chains.

In \cite{kuhn2004coupling}, the authors have established the convergence
of the SAEM-MCMC algorithm to the stationary points. However, their
analysis assumes that complete data likelihood belongs to the curved
exponential family, and all these conditions (M1)-(M5) and (SAEM1)-(SAEM3)
are satisfied. These assumptions are very problem specific, and do
not hold trivially for our case, since our conditional distribution
of the latent variable is extremely complicated. To comment on the
convergence of our proposed algorithm, we need to establish the conditions
(M1)-(M5) and (SAEM1)-(SAEM3) one by one. Finally, we have the convergence
result about our proposed algorithm summarized in the following theorem. 
\begin{thm}
\label{thm:Theorem1}\textup{\emph{Suppose that the}} parameter space\textup{\emph{
$\Theta$ is set to be a sufficiently large bounded set}}\footnote{\emph{This means that the unconstrained maximizer of \eqref{eq:approximated_q}
(given by \eqref{eq:maximizer phi0}, \eqref{eq:maximizer phi 1},
\eqref{eq:maximzer sigma}, and \eqref{eq:maximizer nu}) lies in
this bounded set.}}\textup{\emph{ with the parameter}}\textup{ $\nu>2$}\textup{\emph{,
and the Markov chain generated from \eqref{eq:conditional_distribution_of_tau}
and \eqref{eq:conditional_distribuation_of_y_m} takes values in a
compact set}}\footnote{\emph{Theoretically, the Markov chain generated from \eqref{eq:conditional_distribution_of_tau}
and \eqref{eq:conditional_distribuation_of_y_m} takes its values
in an unbounded set. However, in practice, the chain will not take
very large values, and we can consider the chain takes values in a
very large compact set\cite{kuhn2004coupling,kuhn2005maximum}.}}\textup{\emph{, }}the sequence $\left\{ \boldsymbol{\theta}^{\left(k\right)}\right\} $
generated by Algorithm 1 has the following asymptotic property: with
probability 1, $\lim_{k\rightarrow+\infty}d\left(\boldsymbol{\theta}^{\left(k\right)},\mathcal{L}\right)=0,$
where $d\left(\boldsymbol{\theta}^{\left(k\right)},\mathcal{L}\right)$
denotes the distance from \textup{$\boldsymbol{\theta}^{\left(k\right)}$}
to the set of stationary points of observed data log-likelihood \textup{$\mathcal{L}=\left\{ \boldsymbol{\theta}\in\Theta,\frac{\partial l\left(\boldsymbol{\theta};\mathbf{y}_{\mathsf{o}}\right)}{\partial\boldsymbol{\theta}}=0\right\} $}. 
\end{thm}
\begin{IEEEproof}
Please refer to Appendix \ref{sec:Proof-for-Convergence} for the
proof of the conditions (M1)-(M5) and (SAEM2)-(SAEM3). The condition
(SAEM1) can be be easily satisfied by choosing the step sizes properly
as mentioned before. Upon establishing these conditions, the proof
of this theorem follows straightforward from the analysis of the work
in \cite{kuhn2004coupling}. 
\end{IEEEproof}

\section{Simulations\label{sec:Simulations}}

In this section, we conduct a simulation study of the performance
of the proposed ML estimator and the convergence of the proposed algorithm.
First, we show that the proposed estimator is able to make good estimates
of parameters from the incomplete time series which have been synthesized
to fit the model. Second, we show its robustness to innovation outliers.
Finally, we test it on a real financial time series, the Hang Seng
index.

\subsection{Parameter Estimation}

In this subsection, we show the convergence of the proposed SAEM-MCMC
algorithm and the performance of the proposed estimator on incomplete
Student's $t$ AR(1) time series with different numbers of samples
and missing percentages. The estimation error is measured by the mean
square error (MSE):

\[
\mathsf{MSE}\left(\theta\right)\coloneqq\mathsf{E}\left[\left(\hat{\theta}-\theta^{\mathsf{true}}\right)^{2}\right],
\]
where $\hat{\theta}$ is the estimate for the parameter $\theta$,
and $\theta^{\mathsf{true}}$ is its true value. The parameter $\theta$
can be $\varphi_{0}$, $\varphi_{1}$, $\sigma^{2}$, and $\nu$.
The expectation is approximated via Monte Carlo simulations using
100 independent incomplete time series.

We set $\varphi_{0}^{\mathsf{true}}=1$, $\varphi_{1}^{\mathsf{true}}=0.5$,
$\left(\sigma^{\mathsf{true}}\right)^{2}=0.01$, and $\nu^{\mathsf{true}}=2.5$.
For each incomplete data set $\mathbf{y}_{\mathsf{o}},$ we first
randomly generate a complete time series $\left\{ y_{t}\right\} $
with $T$ samples based on the Student's $t$ AR(1) model. Then $n_{\mathsf{mis}}$
number of samples are randomly deleted to obtain an incomplete time
series. The missing percentage of the incomplete time series is $\rho\coloneqq\frac{n_{\mathsf{mis}}}{T}\times100\%.$

\begin{figure}
\begin{centering}
\textsf{\includegraphics[scale=0.62]{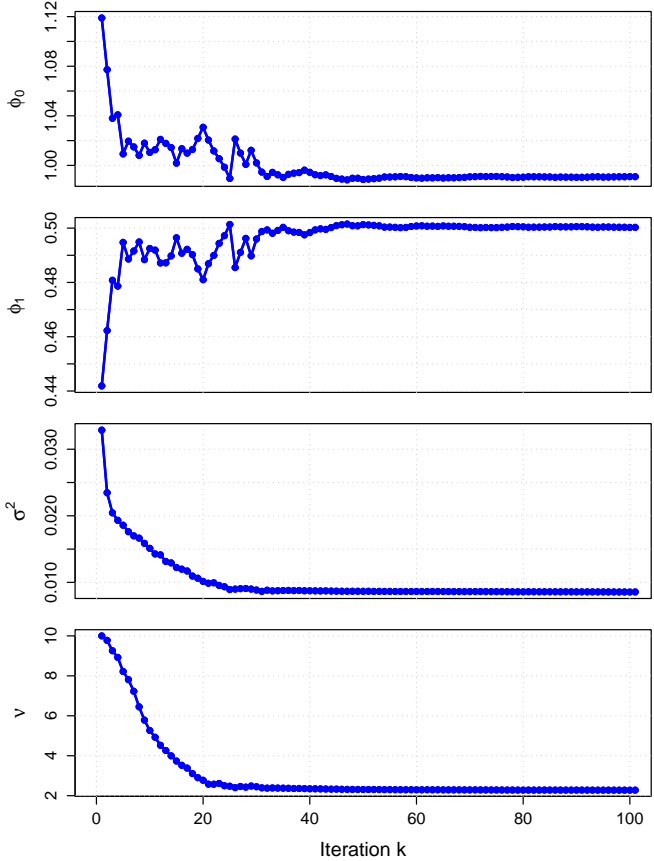}} 
\par\end{centering}
\caption{\label{fig:Estimates-versus-iterations.}Estimates versus iterations.}
\end{figure}

In \prettyref{sec:Convergence}, we have established the convergence
of the proposed SAEM-MCMC algorithm to the stationary points of the
observed data likelihood. However, it is observed that the estimation
result obtained by the algorithm can be sensitive to initializations
due to the existence of multiple stationary points. This is an inevitable
problem since it is a non-convex optimization problem. Interestingly,
it is also observed that when we initialize our algorithm using the
ML estimates assuming the Gaussian AR(1) model, the final estimates
are significantly improved, in comparison to random initializations.
The ML estimation of the Gaussian AR model from incomplete data has
been introduced in \cite{little2002statistical}, and the estimates
can be easily obtained via the deterministic EM algorithm. We initialize
$\varphi_{0}^{\left(0\right)}$, $\varphi_{1}^{\left(0\right)}$,
and $\left(\sigma^{\left(0\right)}\right)^{2}$ use the estimates
from the Gaussian AR(1) model $\left(\varphi_{0}\right)_{\mathsf{g}}$,
$\left(\varphi_{1}\right)_{\mathsf{g}}$, and $\left(\sigma^{2}\right)_{\mathsf{g}}$
, and initialize $\mathbf{y}_{\mathsf{m}}^{\left(0,l\right)}$ using
the mean of the conditional distribution $p\left(\mathbf{y}_{\mathsf{m}};\mathbf{y}_{\mathsf{o}},\left(\varphi_{0}\right)_{\mathsf{g}},\left(\varphi_{1}\right)_{\mathsf{g}},\left(\sigma^{2}\right)_{\mathsf{g}}\right)$,
which is a Gaussian distribution. The parameter $\nu^{\left(0\right)}$
is initialized as a random positive number. In each iteration, we
draw $L=10$ samples. For the step sizes, we set $\gamma^{\left(k\right)}=1$
for 1$\leq k\leq30$ and $\gamma^{\left(k\right)}=\frac{1}{k-30}$
for $k\geq31$. Figure \ref{fig:Estimates-versus-iterations.} gives
an example of applying the proposed SAEM-MCMC algorithm to estimate
the parameters on a synthetic AR(1) data set with $T=300$ and a missing
percentage $\rho=10\%$. We can see that the algorithm converges in
less than 100 iterations, where each iteration just needs $L=10$
runs of Gibbs sampling, and also the final estimation error is small.
Table \ref{tab:Estimation-results-for} compares the estimation results
of the Student's $t$ AR model and the Gaussian AR model. This testifies
our argument that, for incomplete heavy-tailed data, the traditional
method for incomplete Gaussian AR time series is too inefficient,
and significant performance gain can be achieved by designing algorithms
under heavy-tailed model.

\begin{table}[tb]
\caption{\label{tab:Estimation-results-for}Estimation results for incomplete
Student's $t$ AR(1).}

\centering{}%
\begin{tabular}{|c|c|c|c|c|}
\hline 
 & $\hat{\varphi_{0}}$  & $\hat{\varphi_{1}}$  & $\left(\hat{\sigma}\right)^{2}$  & $\hat{\nu}$ \tabularnewline
\hline 
\hline 
True value  & $1$.000  & 0.500  & 0.010  & 2.5\tabularnewline
\hline 
Gaussian AR(1)  & $1.119$  & 0.442  & 0.033  & $+\infty$\tabularnewline
\hline 
Student's $t$ AR(1)  & $0.989$  & 0.501  & 0.009  & $2.234$ \tabularnewline
\hline 
\end{tabular}
\end{table}
\begin{figure}[tbh]
\begin{centering}
\includegraphics[scale=0.62]{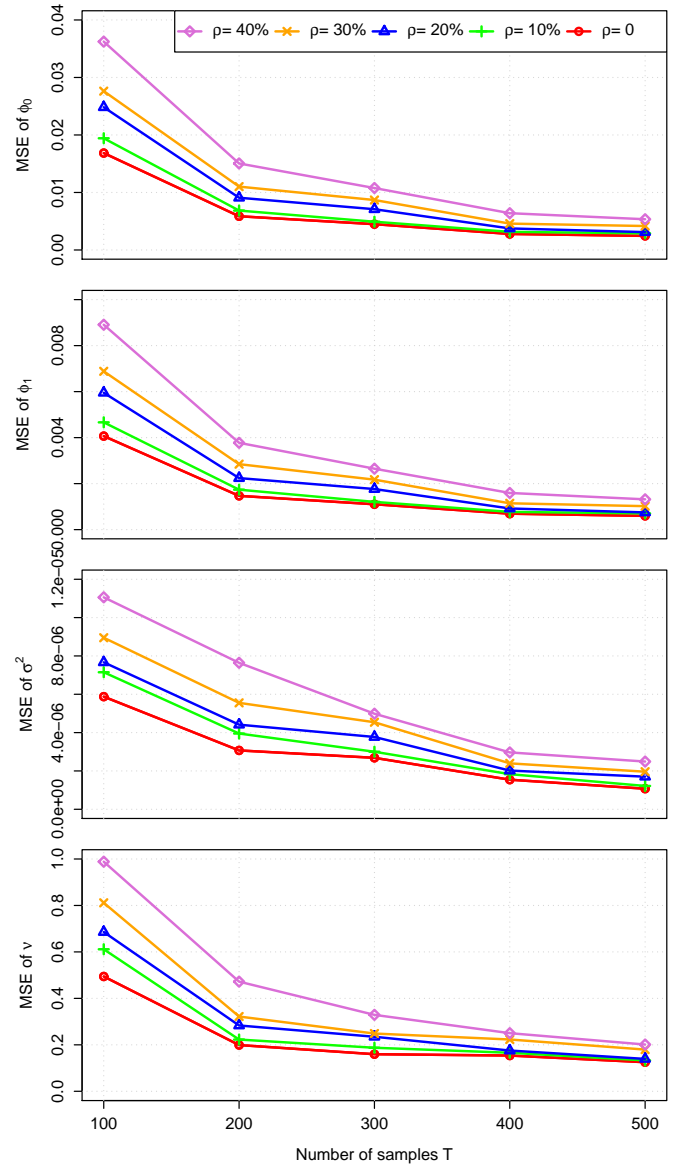} 
\par\end{centering}
\caption{\label{fig:MSEs-dif-number-missing-percentages}MSEs for the incomplete
time series with different number of samples and missing percentages.}
\end{figure}
Figure \ref{fig:MSEs-dif-number-missing-percentages} shows the estimation
results with the numbers of samples $T=100,$ $200$, $300$, $400$,
$500$ and the missing percentages $\rho=$$10\%$, $20\%$, $30\%$,
$40\%$. For reference, we have also given the ML estimation result
from the complete data sets ($\rho=$$0$), which is obtained using
the algorithm in \cite{nduka2018based}. We can observe that our method
performs satisfactorily well even for high percentage of missing data,
and, with increasing sample sizes, the estimates with missing values
match with the estimates of the complete data.

\vspace{-3mm}

\subsection{Robustness to Outliers}

A useful characteristic of the Student's $t$ is its resilience to
outliers, which is not shared by the Gaussian distribution. Here we
illustrate that the Student's $t$ AR model can provide robust estimation
of autoregressive coefficients under innovation outliers.

\begin{figure}[tb]
\begin{centering}
\includegraphics[scale=0.63]{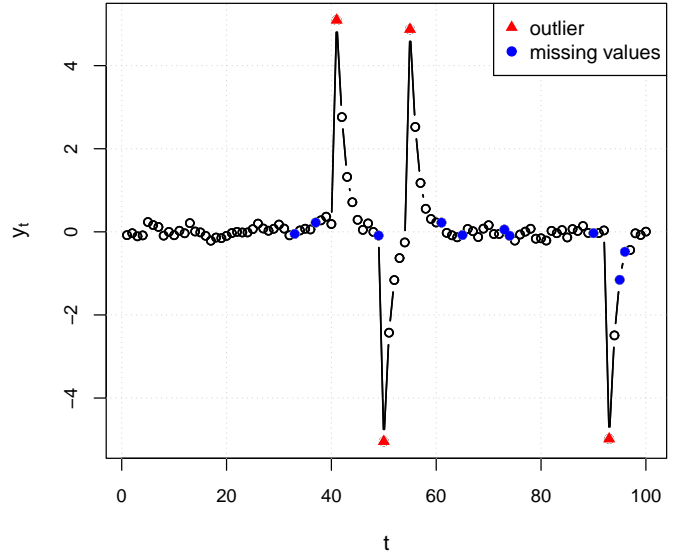} 
\par\end{centering}
\caption{\label{fig:Incomplete-AR(1)-innovation-outliers}Incomplete AR(1)
time series with four innovation outliers.}
\end{figure}

An innovation outlier is an outlier in the $\varepsilon_{t}$ process,
and it is a typical kind of outlier in AR time series \cite{maronna2006timeseries,caroni2004detecting}.
Due to the temporal dependence of AR time series data, an innovation
outlier will affect not only the current observation $y_{t}$, but
also subsequent observations. Figure \ref{fig:Incomplete-AR(1)-innovation-outliers}
gives an example of a Gaussian AR(1) time series contaminated by four
innovation outliers.

When an AR time series is contaminated by outliers, the traditional
ML estimation of autoregressive coefficients based on the Gaussian
AR model, which is equivalent to least squares fitting, will provide
unreliable estimates. Although, for complete time series, there are
numerous works about the robust estimation of autoregressive coefficients
under outliers, unfortunately, less attention was paid to robust estimation
from incomplete time series. As far as we know, only Kharin and Voloshko
have considered robust estimation with missing values \cite{Kharin2011}.
In their paper, they assume that $\phi_{0}$ is known and equal to
$0$. To be consistent with Kharin's method, in this simulation, we
also assume $\varphi_{0}^{\mathsf{true}}$ is known and $\varphi_{0}^{\mathsf{true}}=0$,
although our method can also be applied to the case where $\varphi_{0}^{\mathsf{true}}$
is unknown.

We let $\varphi_{1}^{\mathsf{true}}=0.5$ and $\varepsilon_{t}\overset{i.i.d.}{\sim}\mathcal{N}\left(0,0.01\right)$.
Note here the innovations follow a Gaussian distribution. We randomly
generate an incomplete Gaussian AR(1) time series with $T=100$ samples
and a missing percentage $\rho=0.1$, and it is contaminated by four
innovation outliers. The values of the innovation outliers are set
to be $5$, -5, $5$, $-5$, and the positions are selected randomly.
See Figure \ref{fig:Incomplete-AR(1)-innovation-outliers} for this
incomplete contaminated time series. The Gaussian AR(1) model, the
Student's $t$ AR(1) model, and Kharin's method are applied to estimate
the autoregressive coefficient $\varphi_{1}$. After obtaining the
estimate $\hat{\varphi_{1}}$, we compute the one-step-ahead predictions
$\hat{y}_{t}=\hat{\varphi_{1}}y_{t-1}$ and the prediction error $\left(\hat{y}_{t}-y_{t}\right)^{2}$
for $t\in C_{o}$ and $t-1\in C_{o}$. It is not surprising that the
outliers are poorly predicted, so we omit it when computing the averaged
prediction error. Table \ref{tab:Estimation-and-prediction-Gaussian}
shows the estimation results and the one-step-ahead prediction errors.
It is clear that the ML estimator based on the Gaussian AR(1) has
been significantly affected by the presence of the outliers, while
the Student's $t$ AR(1) model is robust to them, since the outliers
cause the innovations to have a heavy-tailed distribution, which can
be modeled by the Student's $t$ distribution. Kharin's method does
not perform well, either, as this method is designed for addictive
outliers and replacement outliers, rather than innovation outliers.

\begin{table}[tb]
\begin{centering}
\centering{}\caption{\label{tab:Estimation-and-prediction-Gaussian}Estimation and prediction
results for incomplete Gaussian AR(1) time series with outliers.}
\par\end{centering}
\centering{}%
\begin{tabular}{|c|c|c|}
\hline 
 & $\hat{\varphi_{1}}$ ($\varphi_{1}^{\mathsf{true}}=0.5$)  & Averaged prediction error\tabularnewline
\hline 
\hline 
Gaussian AR(1)  & 0.5337  & 0.0121\tabularnewline
\hline 
Student's $t$ AR(1)  & 0.4947  & 0.0110 \tabularnewline
\hline 
Kharin's method  & 0.4210  & 0.0212\tabularnewline
\hline 
\end{tabular}
\end{table}
\begin{table*}[t]
\centering{}\centering{}\caption{\label{tab:Estimation-and-prediction-Hang Seng}Estimation and prediction
results for the Hang Seng index returns.}
\begin{tabular}{|c|c|c|c|c|c|}
\hline 
 & $\hat{\varphi_{0}}$  & $\hat{\varphi_{1}}$  & $\left(\hat{\sigma}\right)^{2}$  & $\hat{\nu}$  & Averaged prediction error\tabularnewline
\hline 
\hline 
Complete data assuming Gaussian innovations  & $7.548\times10^{-4}$  & $-1.058\times10^{-1}$  & $1.702\times10^{-5}$  & $+\infty$  & $9.141\times10^{-6}$\tabularnewline
\hline 
Incomplete data assuming Gaussian innovations  & $8.618\times10^{-4}$  & $-1.253\times10^{-1}$  & $1.665\times10^{-5}$  & $+\infty$  & $9.455\times10^{-6}$\tabularnewline
\hline 
Complete data assuming Student's $t$ innovations  & $5.440\times10^{-4}$  & $-9.580\times10^{-2}$  & $6.524\times10^{-6}$  & $2.622$  & $8.836\times10^{-6}$ \tabularnewline
\hline 
Incomplete data assuming Student's $t$ innovations  & $5.538\times10^{-4}$  & $-9.459\times10^{-2}$  & $6.331\times10^{-6}$  & $2.671$  & $8.831\times10^{-6}$\tabularnewline
\hline 
\end{tabular}
\end{table*}

\subsection{Real Data}

\begin{figure}[tb]
\begin{centering}
\includegraphics[scale=0.5]{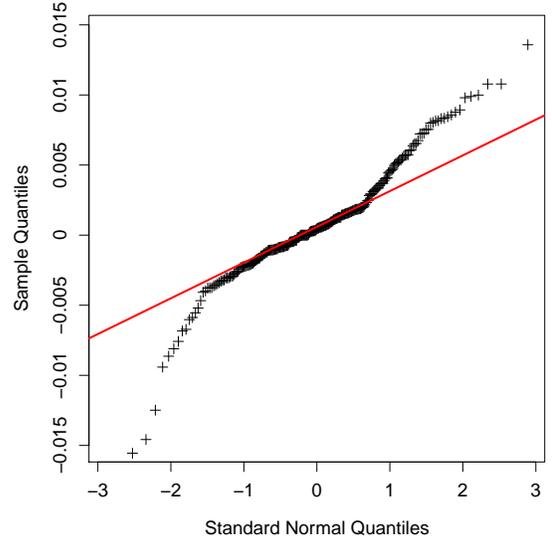} 
\par\end{centering}
\caption{\label{fig:Quantile-quantile-plot-of-Hang-Seng}Quantile-quantile
plot of the the Hang Seng index returns showing that they are heavy-tailed.}
\end{figure}

Here we consider the returns of the Hang Seng index over $260$ working
days from Jan. 2017 to Nov. 2017 (excluding weekends and public holidays).
Figure \ref{fig:Quantile-quantile-plot-of-Hang-Seng} shows the quantile-quantile
(QQ) plot of these returns. The deviation from the straight red line
indicates that the returns are significantly non-Gaussian and heavy-tailed.

We divide the $260$ returns into two parts: the estimation data,
which involves the first $250$ samples, and the test data, which
involves the remaining $10$ samples. First, we fit the estimation
data to the Gaussian AR(1) model and the Student's $t$ AR(1) model,
and estimate the parameters. Then we predict the test data using the
one-step-ahead predication method based on the estimates, and compute
the averaged prediction errors. Next, we randomly delete $10
$ of the estimation data, and estimate the parameters of the Gaussian
AR(1) model and the Student's $t$ AR(1) model from this incomplete
data set. Finally, we also make predictions and compute the averaged
prediction errors based on these estimates of the parameters. The
result is summarized in Table \ref{tab:Estimation-and-prediction-Hang Seng}.
We have the following conclusions: i) the Student's $t$ AR(1) model
performs better than the Gaussian AR(1) model for this heavy-tailed
time series, ii) the proposed parameter estimation method for incomplete
Student's $t$ AR(1) time series can provide similar estimates to
the result of complete data.

\vspace{-3mm}

\section{Conclusions\label{sec:Conclusions}}

In this paper, we have considered parameter estimation of the heavy-tailed
AR model with missing values. We have formulated an ML estimation
problem and developed an efficient approach to obtain the estimates
based on the stochastic EM. Since the conditional distribution of
the latent data in our case is complicated, we proposed a Gibbs sampling
scheme to draw realizations from it. The convergence of the proposed
algorithm to the stationary points has been established. Simulations
show that the proposed approach can provide reliable estimates from
incomplete time series with different percentages of missing values,
and is robust to outliers. Although in this paper we only focus on
the univariate AR model with the Student's $t$ distributed innovations
due to the limit of the space, our method can be extended to multivariate
AR model and also other heavy-tailed distributed innovations.

\vspace{-3mm}

\appendices{}

\section{\textmd{Proof for Lemmas 1 and 2}}

\subsection{\textmd{Proof for \prettyref{lem:lemma1}\label{subsec:Proof-for-Lemma 1}}}

The conditional distribution of $\boldsymbol{\tau}|\mathbf{y}_{\mathsf{m}},\mathbf{y}_{\mathsf{o}};\boldsymbol{\theta}$
is 
\begin{equation}
\begin{aligned} & p\left(\boldsymbol{\tau}|\mathbf{y}_{\mathsf{m}},\mathbf{y}_{\mathsf{o}};\boldsymbol{\theta}\right)\\
 & =\frac{p\left(\mathbf{y},\boldsymbol{\tau};\boldsymbol{\theta}\right)}{p\left(\mathbf{y};\boldsymbol{\theta}\right)}\\
 & \propto p\left(\mathbf{y},\boldsymbol{\tau};\boldsymbol{\theta}\right)\\
 & =\prod_{t=2}^{T}\frac{\left(\frac{\nu}{2}\right)^{\frac{\nu}{2}}\tau_{t}^{\frac{\nu-1}{2}}}{\Gamma\left(\frac{\nu}{2}\right)\sqrt{2\pi\sigma^{2}}}\exp\left(-\frac{\tau_{t}}{2\sigma^{2}}\left(y_{t}-\varphi_{0}-\varphi_{1}y_{t-1}\right)^{2}-\frac{\nu}{2}\tau_{t}\right)\\
 & \propto\prod_{t=2}^{T}\tau_{t}^{\frac{\nu-1}{2}}\exp\left(-\left(\frac{\left(y_{t}-\varphi_{0}-\varphi_{1}y_{t-1}\right)^{2}}{2\sigma^{2}}+\frac{\nu}{2}\right)\tau_{t}\right),
\end{aligned}
\end{equation}
which implies that $\left\{ \tau_{t}\right\} $ are independent from
each other with 
\begin{equation}
\begin{aligned} & p\left(\tau_{t}|\mathbf{y}_{\mathsf{m}},\mathbf{y}_{\mathsf{o}};\boldsymbol{\theta}\right)\\
 & \propto\tau_{t}^{\frac{\nu-1}{2}}\exp\left(-\left(\frac{\left(y_{t}-\varphi_{0}-\varphi_{1}y_{t-1}\right)^{2}}{2\sigma^{2}}+\frac{\nu}{2}\right)\tau_{t}\right).
\end{aligned}
\end{equation}
Comparing this expression with the pdf of the gamma distribution,
we get that $\tau_{t}|\mathbf{y}_{\mathsf{m}},\mathbf{y}_{\mathsf{o}};\boldsymbol{\theta}$
follows a gamma distribution: 
\begin{equation}
\begin{aligned} & \tau_{t}|\mathbf{y}_{\mathsf{m}},\mathbf{y}_{\mathsf{o}};\boldsymbol{\theta}\\
 & \sim Gamma\left(\frac{\nu+1}{2},\thinspace\frac{\left(y_{t}-\varphi_{0}-\varphi_{1}y_{t-1}\right)^{2}/\sigma^{2}+\nu}{2}\right).
\end{aligned}
\end{equation}

\vspace{-3mm}

\subsection{\textmd{Proof for \prettyref{lem:lemma2}\label{subsec:Proof-for-Lemma 2}}}

According to the Gaussian mixture representation \eqref{eq:gaussian mixture presentation of t-1-1}
and \eqref{eq:gaussian mixture presentation of t-2-1}, given $\boldsymbol{\tau}$
and $\boldsymbol{\theta}$, $\varepsilon_{t}$ follows a Gaussian
distribution: $\varepsilon_{t}\overset{i.i.d.}{\sim}\mathcal{N}\left(\mu,\frac{\sigma^{2}}{\tau_{t}}\right)$.
From equation \eqref{eq:ar(1) model}, we can see that, given $\boldsymbol{\tau}$
and $\boldsymbol{\theta}$, the distribution of $y_{t}$ conditional
on all the preceding data $\mathcal{F}_{t-1},$ only depends on the
previous sample $y_{t-1}$: 
\begin{equation}
\begin{aligned}p\left(y_{t}|\boldsymbol{\tau},\mathcal{F}_{t-1};\boldsymbol{\theta}\right)= & p\left(y_{t}|\boldsymbol{\tau},y_{t-1};\boldsymbol{\theta}\right).\end{aligned}
\label{eq:pdf of y_t conditional on previous}
\end{equation}
In addition, the distribution of $y_{t}$ conditional on all the preceding
observed data $\mathcal{F}_{t-1}^{o}$, $\boldsymbol{\tau}$, and
$\boldsymbol{\theta}$, only depends on the nearest observed sample:

\vspace{-3mm}
 
\begin{equation}
\begin{aligned} & p\left(y_{t}|\boldsymbol{\tau},\mathcal{F}_{t-1}^{o};\boldsymbol{\theta}\right)\\
 & =\begin{cases}
p\left(y_{t}|\boldsymbol{\tau},y_{t-1};\boldsymbol{\theta}\right) & t=t_{d}+n_{d}+2,\ldots,t_{d+1},\\
 & for\thinspace d=0,1,\ldots,D,\\
p\left(y_{t}|\boldsymbol{\tau},y_{t-n_{d}-1};\boldsymbol{\theta}\right) & t=t_{d}+n_{d}+1,for\thinspace d=1,2,\ldots,D.
\end{cases}
\end{aligned}
\label{eq:pdf of y_t conditional on previous observed}
\end{equation}
The first case refers to the situation where the previous sample $y_{t-1}$
is observed, while the second case is when $y_{t-1}$ is missing.

Based on the above properties, we have\emph{ }\begin{subequations}
\begin{align}
p\left(\mathbf{y}_{\mathsf{m}}|\boldsymbol{\tau},\mathbf{y}_{\mathsf{o}};\boldsymbol{\theta}\right)= & \frac{\prod_{t=2}^{T}p\left(y_{t}|\boldsymbol{\tau},\mathcal{F}_{t-1};\boldsymbol{\theta}\right)}{\prod_{t\in C_{o}}p\left(y_{t}|\boldsymbol{\tau},\mathcal{F}_{t-1}^{o};\boldsymbol{\theta}\right)}\label{eq:53-b}\\
= & \frac{\prod_{t=2}^{T}p\left(y_{t}|\boldsymbol{\tau},y_{t-1};\boldsymbol{\theta}\right)}{\prod_{d=0}^{D}\prod_{t=t_{d}+n_{d}+2}^{t_{d+1}}p\left(y_{t}|\boldsymbol{\tau},y_{t-1};\boldsymbol{\theta}\right)}\nonumber \\
 & \times\frac{1}{\prod_{d=1}^{D}p\left(y_{t_{d}+n_{d}+1}|\boldsymbol{\tau},y_{t_{d}};\boldsymbol{\theta}\right)}\label{eq:53-d}\\
= & \frac{\prod_{d=1}^{D}\prod_{t=t_{d}+1}^{t_{d}+n_{d}+1}p\left(y_{t}|\boldsymbol{\tau},y_{t-1};\boldsymbol{\theta}\right)}{\prod_{d=1}^{D}p\left(y_{t_{d}+n_{d}+1}|\boldsymbol{\tau},y_{t_{d}};\boldsymbol{\theta}\right)}\label{eq:53-e}\\
= & \prod_{d=1}^{D}\frac{p\left(\mathbf{y}_{d},y_{t_{d}+n_{d}+1}|\boldsymbol{\tau},y_{t_{d}};\boldsymbol{\theta}\right)}{p\left(y_{t_{d}+n_{d}+1}|\boldsymbol{\tau},y_{t_{d}};\boldsymbol{\theta}\right)}\label{eq:53-f}\\
= & \prod_{d=1}^{D}p\left(\mathbf{y}_{d}|\boldsymbol{\tau},y_{t_{d}},y_{t_{d}+n_{d}+1};\boldsymbol{\theta}\right),\label{eq:53-g}
\end{align}
\end{subequations}where the equations \eqref{eq:53-b} and \eqref{eq:53-g}
are from the definition of conditional pdf, the equation \eqref{eq:53-d}
is from \eqref{eq:pdf of y_t conditional on previous} and \eqref{eq:pdf of y_t conditional on previous observed}.
The equation \eqref{eq:53-g} implies that the different missing blocks
$\left\{ \mathbf{y}_{d}\right\} $ are independent from each other,
and the conditional distribution of $\mathbf{y}_{d}$ only depends
on the two nearest observed samples $y_{t_{d}}$ and $y_{t_{d}+n_{d}+1}.$

To obtain the pdf of the missing block $p\left(\mathbf{y}_{d}|\boldsymbol{\tau},y_{t_{d}},y_{t_{d}+n_{d}+1};\boldsymbol{\theta}\right)$,
we first analyze the joint pdf of the missing block and next observed
sample $\mathbf{y}_{cd}=\left[\mathbf{y}_{d}^{T},\thinspace y_{t_{d}+n_{d}+1}\right]^{T}=\left[y_{t_{d}+1},y_{t_{d}+2},\ldots,y_{t_{d}+n_{d}+1}\right]$:
$p\left(\mathbf{y}_{cd}|\boldsymbol{\tau},y_{t_{d}};\boldsymbol{\theta}\right)$.
Given $\boldsymbol{\tau}$, $y_{t_{d}}$, and $\boldsymbol{\theta}$,
from \eqref{eq:ar(1) model}, we have

\vspace{-3mm}

\begin{equation}
\begin{aligned}y_{t_{d}+i}= & \varphi_{0}+\varphi_{1}y_{t_{d}+i-1}+\varepsilon_{t_{d}+i}\\
= & \varphi_{0}+\varphi_{1}\left(\varphi_{0}+\varphi_{1}y_{t_{d}+i-2}+\varepsilon_{t_{d}+i-1}\right)+\varepsilon_{t_{d}+i}\\
= & \varphi_{0}+\varphi_{1}\varphi_{0}+\varphi_{1}^{2}y_{t_{d}+i-2}+\varphi_{1}\varepsilon_{t_{d}+i-1}+\varepsilon_{t_{d}+i}\\
= & \sum_{q=0}^{i-1}\varphi_{1}^{q}\varphi_{0}+\varphi_{1}^{i}y_{t_{d}}+\sum_{q=1}^{i}\varphi_{1}^{\left(i-q\right)}\varepsilon_{t_{d}+q},
\end{aligned}
\end{equation}
for $i=1,2,\ldots$,$n_{d}+1$, which means that $y_{t_{d}+i}$ can
be expressed as the sum of the constant $\sum_{q=0}^{i-1}\varphi_{1}^{q}\varphi_{0}+\varphi_{1}^{i}y_{t_{d}}$
and a linear combination of the independent Gaussian random variables
$\varepsilon_{t_{d}+1},$ $\varepsilon_{t_{d}+2}$, $\ldots$, $\varepsilon_{t_{d}+i}.$
Therefore, we can obtain that $\mathbf{y}_{cd}$ follows a Gaussian
distribution as follows: 
\begin{equation}
\mathbf{y}_{cd}|\boldsymbol{\tau},y_{t_{d}};\boldsymbol{\theta}\sim\mathcal{N}\left(\boldsymbol{\mu}_{cd},\boldsymbol{\Sigma}_{cd}\right),\label{eq:conditional distribution}
\end{equation}
where the $i$-th component of $\boldsymbol{\mu}_{cd}$ 
\begin{equation}
\begin{aligned}\mu_{cd(i)}= & \mathsf{E}\left[y_{t_{d}+i}\right]\\
= & \mathsf{E}\left[\sum_{q=0}^{i-1}\varphi_{1}^{q}\varphi_{0}+\varphi_{1}^{i}y_{t_{d}}+\sum_{q=1}^{i}\varphi_{1}^{\left(i-q\right)}\varepsilon_{t_{d}+q}\right]\\
= & \sum_{q=0}^{i-1}\varphi_{1}^{q}\varphi_{0}+\varphi_{1}^{i}y_{t_{d}}+\sum_{q=1}^{i}\varphi_{1}^{\left(i-q\right)}\mathsf{E}\left[\varepsilon_{t_{d}+q}\right]\\
= & \sum_{q=0}^{i-1}\varphi_{1}^{q}\varphi_{0}+\varphi_{1}^{i}y_{t_{d}},
\end{aligned}
\label{eq:mean of cd}
\end{equation}
and the component in the $i$-th column and the $j$-th row of $\boldsymbol{\Sigma}_{cd}$
\begin{equation}
\begin{aligned}\Sigma_{cd(i,j)}= & \mathsf{E}\left[\left(y_{t_{d}+i}-\mu_{cd(i)}\right)\left(y_{t_{d}+j}-\mu_{cd(j)}\right)\right]\\
= & \mathsf{E}\left[\left(\sum_{q_{1}=1}^{i}\varphi_{1}^{\left(i-q_{1}\right)}\varepsilon_{t_{d}+q_{1}}\right)\left(\sum_{q_{2}=1}^{j}\varphi_{1}^{\left(j-q_{2}\right)}\varepsilon_{t_{d}+q_{2}}\right)\right]\\
= & \sum_{q_{1}=1}^{i}\sum_{q_{2}=1}^{j}\varphi_{1}^{\left(i+j-q_{1}-q_{2}\right)}\mathsf{E}\left[\varepsilon_{t_{d}+q_{1}}\varepsilon_{t_{d}+q_{2}}\right]\\
= & \sigma^{2}\sum_{q=1}^{\min\left(i,j\right)}\frac{\varphi_{1}^{\left(i+j-2q\right)}}{\tau_{t_{d}+q}}.
\end{aligned}
\label{eq:covariance of cd}
\end{equation}
with the last equation following from 
\[
\mathsf{E}\left[\varepsilon_{t_{d}+q_{1}}\varepsilon_{t_{d}+q_{2}}\right]=\begin{cases}
\frac{\sigma^{2}}{\tau_{t_{d}+q_{1}}}, & q_{1}=q_{2};\\
0, & q_{1}\neq q_{2}.
\end{cases}
\]

Recall that $p\left(\mathbf{y}_{d}|\boldsymbol{\tau},y_{t_{d}},y_{t_{d}+n_{d}+1};\boldsymbol{\theta}\right)$
is a conditional pdf of $p\left(\mathbf{y}_{d},y_{t_{d}+n_{d}+1}|\boldsymbol{\tau},y_{t_{d}};\boldsymbol{\theta}\right)$.
Since conditional distributions of a Gaussian distribution is Gaussian,
we can get that $\mathbf{y}_{d}|\boldsymbol{\tau},y_{t_{d}},y_{t_{d}+n_{d}+1};\boldsymbol{\theta}$
follows a Gaussian distribution as \eqref{eq:distribution of missing block}.
The parameters of this conditional distribution can be computed based
on \vspace{-3mm}
 
\begin{equation}
\mathbf{\boldsymbol{\mu}}_{d}=\boldsymbol{\mu}_{cd(1:n_{d})}+\frac{\boldsymbol{\Sigma}_{cd(1:n_{d},n_{d}+1)}}{\Sigma_{cd(n_{d}+1,n_{d}+1)}}\left(y_{t_{d}+n_{d}+1}-\mu_{cd(n_{d}+1)}\right),\label{eq:conditional mean of missing block}
\end{equation}
and\vspace{-3mm}
 
\begin{equation}
\boldsymbol{\Sigma}_{d}=\boldsymbol{\Sigma}_{cd(1:n_{d},1:n_{d})}-\frac{\boldsymbol{\Sigma}_{cd(1:n_{d},n_{d}+1)}\boldsymbol{\Sigma}_{cd(n_{d}+1,1:n_{d})}}{\Sigma_{cd(n_{d}+1,n_{d}+1)}},\label{eq:conditional covariance of missing block}
\end{equation}
where $\boldsymbol{\mu}_{cd(a_{1}:a_{2})}$ denotes the subvector
consisting of the $a_{1}$-th to $a_{2}$-th component of $\mathbf{\boldsymbol{\mu}}_{cd}$,
and the $\boldsymbol{\Sigma}_{cd(a_{1}:a_{2},b_{1}:b_{2})}$ means
the submatrix consisting of the components in the $a_{1}$-th to $a_{2}$-th
rows and the $b1$-th to $b_{2}$-th columns of $\boldsymbol{\Sigma}_{cd}$.
Plugging the equations \eqref{eq:mean of cd} and \eqref{eq:covariance of cd}
into the equations \eqref{eq:conditional mean of missing block} and
\eqref{eq:conditional covariance of missing block} gives the equations
\eqref{eq:mean of missing block} and \eqref{eq:covariance of missing block},
respectively.

\vspace{-3mm}

\section{\textmd{Proof for Conditions (M1)-(M5) and (SAEM2)-(SAEM3)\label{sec:Proof-for-Convergence}}}

In this section, we will establish the listed conditions one by one.
The observed data $\mathbf{y}_{\mathsf{o}}$ is known. We assume that
$\mathbf{y}_{\mathsf{o}}$ is finite. Since the parameter space $\Theta$
is a large bounded set with $\nu>2$, we can assume that $|\varphi_{0}|<\varphi_{0}^{+},|\varphi_{1}|<\varphi_{1}^{+},\sigma>\sigma^{-},$
and $\nu^{-}<\nu<\nu^{+}$, where $\varphi_{0}^{+}$, $\varphi_{1}^{+}$,
and $\nu^{+}$ are very large positive numbers, $\sigma^{-}$ is a
very small positive number, and $\nu^{-}$ is a very small positive
number satisfying $\nu^{-}\geq2$.{} We first prove the conditions
(M1)-(M5), then prove the conditions (SAEM2) and (SAEM3). 

\vspace{-2mm}

\subsection{\textmd{Proof of (M1)-(M5)}}

The proof begins by establishing the following two intermediary lemmas.

\vspace{-2mm}

\begin{lem}
\textup{\emph{\label{lem:lemma3}For any $\mathbf{y}_{\mathsf{o}}$
and }}\textup{$\boldsymbol{\theta}\in\Theta$, $p\left(\mathbf{y}_{\mathsf{o}};\boldsymbol{\theta}\right)=\int\int p\left(\mathbf{y},\boldsymbol{\tau};\boldsymbol{\theta}\right)\mathsf{d}\mathbf{y}_{\mathsf{m}}\mathsf{d\boldsymbol{\tau}}=\int p\left(\mathbf{y};\boldsymbol{\theta}\right)\mathsf{d}\mathbf{y}_{\mathsf{m}}<\infty.$} 
\end{lem}
\begin{lem}
\textup{\emph{\label{lem:lemma4}For any $\mathbf{y}_{\mathsf{o}}$
, $\boldsymbol{\theta}\in\Theta$ and $1<t\leq T$ 
\begin{equation}
\iint g\left(\mathbf{y},\boldsymbol{\tau}\right)p\left(\mathbf{y},\boldsymbol{\tau};\boldsymbol{\theta}\right)\mathsf{d}\mathbf{y}_{\mathsf{m}}\mathsf{d\boldsymbol{\tau}}<\infty,
\end{equation}
where $g\left(\mathbf{y},\boldsymbol{\tau}\right)$ can be $\tau_{t}$,
$\tau_{t}^{2}$, $y_{t}^{2}$,$\tau_{t}y_{t-1}^{2}$, $\tau_{t}y_{t}^{2}$,
or }}\textup{$-\log\left(\tau_{t}\right)$} 
\end{lem}
\prettyref{lem:lemma3} indicates that the observed data likelihood
$p\left(\mathbf{y}_{\mathsf{o}};\boldsymbol{\theta}\right)$ is bounded,
and \prettyref{lem:lemma4} shows that the expectation of $g\left(\mathbf{y},\boldsymbol{\tau}\right)$
is bounded. These lemmas provide the key ingredients required for
establishing (M1)-(M5), and their usage for subsequent analysis is
self-explanatory. Due to space limitations, we do not include their
proofs here. Interested readers may refer to the supplementary material.

(M1) For condition (M1), based on \eqref{eq:minimal sufficient statistics},
we can get

\vspace{-2mm}

\begin{equation}
\begin{aligned} & \int\int\Vert\mathbf{s}\left(\mathbf{y}_{\mathsf{o}},\mathbf{y}_{\mathsf{m}},\boldsymbol{\tau}\right)\Vert p\left(\mathbf{y}_{\mathsf{m}},\boldsymbol{\tau}|\mathbf{y}_{\mathsf{o}};\boldsymbol{\theta}\right)\mathsf{d}\mathbf{y}_{\mathsf{m}}\mathsf{d\boldsymbol{\tau}}\\
 & =\frac{\int\int\Vert\mathbf{s}\left(\mathbf{y}_{\mathsf{o}},\mathbf{y}_{\mathsf{m}},\boldsymbol{\tau}\right)\Vert p\left(\mathbf{y}_{\mathsf{o}},\mathbf{y}_{\mathsf{m}},\boldsymbol{\tau};\boldsymbol{\theta}\right)\mathsf{d}\mathbf{y}_{\mathsf{m}}\mathsf{d\boldsymbol{\tau}}}{p\left(\mathbf{y}_{\mathsf{o}};\boldsymbol{\theta}\right)}\\
 & \leq\frac{1}{p\left(\mathbf{y}_{\mathsf{o}};\boldsymbol{\theta}\right)}\sum_{t=2}^{T}\int\int\biggl(\bigr|\log\left(\tau_{t}\right)-\tau_{t}\bigr|+\bigr|\tau_{t}y_{t}^{2}\bigr|+\bigr|\tau_{t}\bigr|\\
 & \hspace{9.5em}+\bigr|\tau_{t}y_{t-1}^{2}\bigr|+\bigr|\tau_{t}y_{t}\bigr|+\bigr|\tau_{t}y_{t}y_{t-1}\bigr|\\
 & \hspace{9.5em}+\bigr|\tau_{t}y_{t-1}\bigr|\biggr)p\left(\mathbf{y}_{\mathsf{o}},\mathbf{y}_{\mathsf{m}},\boldsymbol{\tau};\boldsymbol{\theta}\right)\mathsf{d}\mathbf{y}_{\mathsf{m}}\mathsf{d\boldsymbol{\tau}}\\
 & \leq\frac{1}{p\left(\mathbf{y}_{\mathsf{o}};\boldsymbol{\theta}\right)}\sum_{t=2}^{T}\int\int\biggl(\tau_{t}-\log\left(\tau_{t}\right)+\tau_{t}y_{t}^{2}+\tau_{t}\\
 & \hspace{9.5em}+\tau_{t}y_{t-1}^{2}+\frac{\tau_{t}^{2}+y_{t}^{2}}{2}+\frac{\tau_{t}\left(y_{t}^{2}+y_{t-1}^{2}\right)}{2}\\
 & \hspace{9.5em}+\frac{\tau_{t}^{2}+y_{t-1}^{2}}{2}\biggr)p\left(\mathbf{y}_{\mathsf{o}},\mathbf{y}_{\mathsf{m}},\boldsymbol{\tau};\boldsymbol{\theta}\right)\mathsf{d}\mathbf{y}_{\mathsf{m}}\mathsf{d\boldsymbol{\tau}}\\
 & <\infty,
\end{aligned}
\end{equation}
where the three inequalities follow from the triangular inequality,
the property of squares $x_{1}x_{2}\leq\frac{x_{1}^{2}+x_{2}^{2}}{2}$,
and \prettyref{lem:lemma4}, respectively.

(M2) From the definition of $\psi\left(\boldsymbol{\theta}\right)$
and $\boldsymbol{\phi}\left(\boldsymbol{\theta}\right)$ in \eqref{eq:psi}
and \eqref{eq:phi}, their continuous differentiability can be easily
verified.

(M3) For condition (M3), 
\begin{equation}
\begin{alignedat}{1}\bar{\mathbf{s}}\left(\boldsymbol{\theta}\right)= & \int\int\mathbf{s}\left(\mathbf{y}_{\mathsf{o}},\mathbf{y}_{\mathsf{m}},\boldsymbol{\tau}\right)p\left(\mathbf{y}_{\mathsf{m}},\boldsymbol{\tau}|\mathbf{y}_{\mathsf{o}};\boldsymbol{\theta}\right)\mathsf{d}\mathbf{y}_{\mathsf{m}}\mathsf{d\boldsymbol{\tau}}\\
= & \int\int\mathbf{s}\left(\mathbf{y}_{\mathsf{o}},\mathbf{y}_{\mathsf{m}},\boldsymbol{\tau}\right)\frac{p\left(\mathbf{y},\boldsymbol{\tau};\boldsymbol{\theta}\right)}{p\left(\mathbf{y}_{\mathsf{o}};\boldsymbol{\theta}\right)}\mathsf{d}\mathbf{y}_{\mathsf{m}}\mathsf{d\boldsymbol{\tau}}\\
= & \frac{\int\int\mathbf{s}\left(\mathbf{y}_{\mathsf{o}},\mathbf{y}_{\mathsf{m}},\boldsymbol{\tau}\right)p\left(\mathbf{y},\boldsymbol{\tau};\boldsymbol{\theta}\right)\mathsf{d}\mathbf{y}_{\mathsf{m}}\mathsf{d\boldsymbol{\tau}}}{\int\int p\left(\mathbf{y},\boldsymbol{\tau};\boldsymbol{\theta}\right)\mathsf{d}\mathbf{y}_{\mathsf{m}}\mathsf{d\boldsymbol{\tau}}}.
\end{alignedat}
\end{equation}
Since $\int\int p\left(\mathbf{y},\boldsymbol{\tau};\boldsymbol{\theta}\right)\mathsf{d}\mathbf{y}_{\mathsf{m}}\mathsf{d\boldsymbol{\tau}}=p\left(\mathbf{y}_{\mathsf{o}};\boldsymbol{\theta}\right)>0$
and $p\left(\mathbf{y},\boldsymbol{\tau};\boldsymbol{\theta}\right)$
is continuously differentiable, which can be easily checked from its
definition \eqref{eq:expectation of complete data log-likelihood},
we can get that $\bar{\mathbf{s}}\left(\boldsymbol{\theta}\right)$
is continuously differentiable.

(M4) Since $\int\int p\left(\mathbf{y},\boldsymbol{\tau};\boldsymbol{\theta}\right)\mathsf{d}\mathbf{y}_{\mathsf{m}}\mathsf{d\boldsymbol{\tau}}>0$,
and $p\left(\mathbf{y},\boldsymbol{\tau};\boldsymbol{\theta}\right)$
is 7 times differentiable, $l\left(\boldsymbol{\theta};\mathbf{y}_{\mathsf{o}}\right)=\log\left(\int\int p\left(\mathbf{y},\boldsymbol{\tau};\boldsymbol{\theta}\right)\mathsf{d}\mathbf{y}_{\mathsf{m}}\mathsf{d\boldsymbol{\tau}}\right)$
is 7 times differentiable. For the verification of the equation \eqref{eq:change the order of derivative and integral},
according to Leibniz integral rule, the equation \eqref{eq:change the order of derivative and integral}
holds under the following three conditions: 
\begin{enumerate}
\item $\int\int p\left(\mathbf{y},\boldsymbol{\tau};\boldsymbol{\theta}\right)\mathsf{d}\mathbf{y}_{\mathsf{m}}\mathsf{d\boldsymbol{\tau}}<\infty$, 
\item $\frac{\partial p\left(\mathbf{y},\boldsymbol{\tau};\boldsymbol{\theta}\right)}{\partial\boldsymbol{\theta}}$
exists for all the $\boldsymbol{\theta}\in\Theta$, 
\item there is an integrable function $g\left(\mathbf{y},\boldsymbol{\tau}\right)$
such that $\Bigl|$$\frac{\partial p\left(\mathbf{y},\boldsymbol{\tau};\boldsymbol{\theta}\right)}{\partial\boldsymbol{\theta}}\Bigr|\leq g\left(\mathbf{y},\boldsymbol{\tau}\right)$
for all $\boldsymbol{\theta}\in\Theta$ and almost every $\mathbf{y}$
and $\boldsymbol{\tau}$. 
\end{enumerate}
Since the first condition has been proved in \prettyref{lem:lemma3},
and the second condition can be easily verified from its definition,
here we focus on the third condition.

From the equation \eqref{eq:complete data likelihood for ar(1)},
the derivative of $p\left(\mathbf{y},\boldsymbol{\tau};\boldsymbol{\theta}\right)$
with respect to $\varphi_{0}$ is

\vspace{-2mm}
 
\begin{equation}
\begin{aligned} & \biggl|\frac{\partial p\left(\mathbf{y},\boldsymbol{\tau};\boldsymbol{\theta}\right)}{\partial\varphi_{0}}\biggl|\\
 & =\Biggl|p\left(\mathbf{y},\boldsymbol{\tau};\boldsymbol{\theta}\right)\sum_{j=2}^{T}\frac{\tau_{j}\left(y_{j}-\varphi_{0}-\varphi_{1}y_{j-1}\right)}{\sigma^{2}}\Biggr|\\
 & \leq\frac{p\left(\mathbf{y},\boldsymbol{\tau};\boldsymbol{\theta}\right)}{\sigma^{2}}\sum_{j=2}^{T}\left(|\tau_{j}y_{j}|+|\varphi_{0}\tau_{j}|+|\varphi_{1}\tau_{j}y_{j-1}|\right)\\
 & \leq\frac{p\left(\mathbf{y},\boldsymbol{\tau};\boldsymbol{\theta}^{\ast}\right)}{\left(\sigma^{-}\right)^{2}}\sum_{j=2}^{T}\biggl\{\biggl(\frac{\tau_{j}^{2}+y_{j}^{2}}{2}+\varphi_{0}^{+}\tau_{j}+\frac{\varphi_{1}^{+}\left(y_{j-1}^{2}+\tau_{j}^{2}\right)}{2}\biggr)\biggr\}\\
 & =g_{\varphi_{0}}\left(\mathbf{y},\boldsymbol{\tau}\right),
\end{aligned}
\end{equation}
where $\boldsymbol{\theta}^{*}=\underset{\boldsymbol{\theta}\in\Theta}{\arg\max\ }p\left(\mathbf{y},\boldsymbol{\tau};\boldsymbol{\theta}\right).$
The first inequality follows from the triangle inequality, and the
second inequality follows from $p\left(\mathbf{y},\boldsymbol{\tau};\boldsymbol{\theta}^{\ast}\right)\geq p\left(\mathbf{y},\boldsymbol{\tau};\boldsymbol{\theta}\right)$,
$|\varphi_{0}|<\varphi_{0}^{+}$, $|\varphi_{1}|<\varphi_{1}^{+}$,
$\sigma>\sigma^{-}$, and the property of squares.

The derivative with respect to $\varphi_{1}$ is

\vspace{-3mm}

\begin{align*}
 & \biggl|\frac{\partial p\left(\mathbf{y},\boldsymbol{\tau};\boldsymbol{\theta}\right)}{\partial\varphi_{1}}\biggl|\\
 & =\biggr|p\left(\mathbf{y},\boldsymbol{\tau};\boldsymbol{\theta}\right)\sum_{j=2}^{T}\frac{1}{\sigma^{2}}\tau_{j}y_{j-1}\left(y_{j}-\varphi_{0}-\varphi_{1}y_{j-1}\right)\biggr|\hspace{4em}
\end{align*}

\begin{align}
 & \leq\frac{p\left(\mathbf{y},\boldsymbol{\tau};\boldsymbol{\theta}\right)}{\sigma^{2}}\sum_{j=2}^{T}\biggl(|\tau_{j}y_{j}y_{j-1}|+|\varphi_{0}\tau_{j}y_{j-1}|+|\varphi_{1}\tau_{j}y_{j-1}^{2}|\biggr)\nonumber \\
 & \leq\frac{p\left(\mathbf{y},\boldsymbol{\tau};\boldsymbol{\theta}^{\ast}\right)}{\left(\sigma^{-}\right)^{2}}\sum_{j=2}^{T}\biggl(\frac{\tau_{j}\left(y_{j}^{2}+y_{j-1}^{2}\right)}{2}+\frac{\varphi_{0}^{+}\left(\tau_{j}^{2}+y_{j-1}^{2}\right)}{2}\nonumber \\
 & \hspace{8.5em}+\varphi_{1}^{+}\tau_{j}y_{j-1}^{2}\biggr)\nonumber \\
 & =g_{\varphi_{1}}\left(\mathbf{y},\boldsymbol{\tau}\right),
\end{align}
where the first inequality follows from the triangle inequality, and
the second inequality follows from $|\varphi_{0}|<\varphi_{0}^{+}$,
$|\varphi_{1}|<\varphi_{1}^{+}$, $\sigma>\sigma^{-}$, and the property
of squares.

The derivative with respect to $\sigma^{2}$ is

\vspace{-3mm}
 
\begin{equation}
\begin{aligned} & \biggl|\frac{\partial p\left(\mathbf{y},\boldsymbol{\tau};\boldsymbol{\theta}\right)}{\partial\sigma^{2}}\biggl|\\
 & =p\left(\mathbf{y},\boldsymbol{\tau};\boldsymbol{\theta}\right)\sum_{j=2}^{T}\left\{ \frac{\tau_{j}}{2\sigma^{4}}\left(y_{j}-\varphi_{0}-\varphi_{1}y_{j-1}\right)^{2}-\frac{1}{2\sigma^{2}}\right\} \biggr|\\
 & \leq p\left(\mathbf{y},\boldsymbol{\tau};\boldsymbol{\theta}\right)\sum_{j=2}^{T}\left\{ \frac{\tau_{j}}{2\sigma^{4}}\left(y_{j}-\varphi_{0}-\varphi_{1}y_{j-1}\right)^{2}+\frac{1}{2\sigma^{2}}\right\} \\
 & \leq p\left(\mathbf{y},\boldsymbol{\tau};\boldsymbol{\theta}\right)\sum_{j=2}^{T}\left\{ \frac{\tau_{j}}{2\sigma^{4}}\left(2\left(y_{j}-\varphi_{0}\right)^{2}+2\varphi_{1}^{2}y_{j-1}^{2}\right)+\frac{1}{2\sigma^{2}}\right\} \\
 & \leq p\left(\mathbf{y},\boldsymbol{\tau};\boldsymbol{\theta}\right)\sum_{j=2}^{T}\left\{ \frac{\tau_{j}}{2\sigma^{4}}\left(4y_{j}^{2}+4\varphi_{0}^{2}+2\varphi_{1}^{2}y_{j-1}^{2}\right)+\frac{1}{2\sigma^{2}}\right\} \\
 & \leq p\left(\mathbf{y},\boldsymbol{\tau};\boldsymbol{\theta}^{\ast}\right)\sum_{j=2}^{T}\biggl\{\frac{\tau_{j}}{2\left(\sigma^{-}\right)^{2}}\left(4y_{j}^{2}+4\left(\varphi_{0}^{+}\right)^{2}+2\left(\varphi_{1}^{+}\right)^{2}y_{j-1}^{2}\right)\\
 & \hspace{7em}+\frac{1}{2\left(\sigma^{-}\right)^{2}}\biggr\}\\
 & =g_{\sigma^{2}}\left(\mathbf{y},\boldsymbol{\tau}\right),
\end{aligned}
\end{equation}
where the first inequality follows from the triangle inequality, the
second and third inequalities follow from the property of squares
$\left(x_{1}-x_{2}\right)^{2}\leq2\left(x_{1}^{2}+x_{2}^{2}\right)$,
and the last inequality follows from $p\left(\mathbf{y},\boldsymbol{\tau};\boldsymbol{\theta}^{\ast}\right)\geq p\left(\mathbf{y},\boldsymbol{\tau};\boldsymbol{\theta}\right)$,
$|\varphi_{0}|<\varphi_{0}^{+}$, $|\varphi_{1}|<\varphi_{1}^{+}$,
and $\sigma>\sigma^{-}$.

The derivative with respect to $\nu$ is 
\begin{equation}
\begin{aligned} & \biggl|\frac{\partial p\left(\mathbf{y},\boldsymbol{\tau};\boldsymbol{\theta}\right)}{\partial\nu}\biggl|\\
 & =\biggr|p\left(\mathbf{y},\boldsymbol{\tau};\boldsymbol{\theta}\right)\sum_{j=2}^{T}\frac{1}{2}\left(1+\log\left(\frac{\nu}{2}\right)-\Psi\left(\frac{\nu}{2}\right)+\log\left(\tau_{j}\right)-\tau_{j}\right)\biggr|\\
 & \leq\frac{1}{2}p\left(\mathbf{y},\boldsymbol{\tau};\boldsymbol{\theta}\right)\sum_{j=2}^{T}\left\{ \biggr|1+\log\left(\frac{\nu}{2}\right)-\Psi\left(\frac{\nu}{2}\right)\biggr|+\biggr|\log\left(\tau_{j}\right)-\tau_{j}\biggr|\right\} \\
 & \leq p\left(\mathbf{y},\boldsymbol{\tau};\boldsymbol{\theta}^{\ast}\right)\sum_{j=2}^{T}\biggl(\frac{1}{2}+\frac{1}{2}\log\left(\frac{\nu^{-}}{2}\right)-\frac{1}{2}\Psi\left(\frac{\nu^{-}}{2}\right)\\
 & \hspace{7em}+\frac{1}{2}\tau_{j}-\frac{1}{2}\log\left(\tau_{j}\right)\biggr)\\
 & =g_{\nu}\left(\mathbf{y},\boldsymbol{\tau}\right),
\end{aligned}
\end{equation}
where $\varPsi\left(\cdot\right)$ is the digamma function. The first
inequality follows from the triangle inequality, and the second inequality
is due to that $\log\left(\frac{\nu}{2}\right)-\varPsi\left(\frac{\nu}{2}\right)$
is positive and strictly decreasing for $\nu\geq\nu^{-}$\cite{liu1995ml}.

Based on Lemmas \ref{lem:lemma3} and \ref{lem:lemma4}, we can obtain
that $\iint g_{\varphi_{0}}\left(\mathbf{y},\boldsymbol{\tau},\right)\mathsf{d}\mathbf{y}_{\mathsf{m}}\mathsf{d\boldsymbol{\tau}}<\infty,$
$\iint g_{\varphi_{1}}\left(\mathbf{y},\boldsymbol{\tau}\right)\mathsf{d}\mathbf{y}_{\mathsf{m}}\mathsf{d\boldsymbol{\tau}}<\infty$,
$\iint g_{\sigma^{2}}\left(\mathbf{y},\boldsymbol{\tau}\right)\mathsf{d}\mathbf{y}_{\mathsf{m}}\mathsf{d\boldsymbol{\tau}}<\infty$,
and $\iint g_{\nu}\left(\mathbf{y},\boldsymbol{\tau}\right)\mathsf{d}\mathbf{y}_{\mathsf{m}}\mathsf{d\boldsymbol{\tau}}<\infty.$
The condition (M4) is verified.

(M5) This condition requires the existence of the global maximizer
$\tilde{\boldsymbol{\theta}}\left(\bar{\mathbf{s}}\right)$ for $Q\left(\boldsymbol{\theta},\bar{\mathbf{s}}\right)$
and its continuous differentiability. Since $Q\left(\boldsymbol{\theta},\bar{\mathbf{s}}\right)$
takes the same form with $\hat{Q}\left(\boldsymbol{\theta},\hat{\mathbf{s}}^{\left(k\right)}\right),$
the maximizer will also take the same form. From \eqref{eq:maximizer phi0}-\eqref{eq:maximizer nu},
we have 
\begin{equation}
\tilde{\varphi_{0}}\left(\bar{\mathbf{s}}\right)=\frac{\bar{s}_{5}-\tilde{\varphi_{1}}\left(\bar{\mathbf{s}}\right)\bar{s}_{7}}{\bar{s}_{3}},\label{eq:maximizer phi0-1-1}
\end{equation}
\begin{equation}
\tilde{\varphi_{1}}\left(\bar{\mathbf{s}}\right)=\frac{\bar{s}_{3}\bar{s}_{6}-\bar{s}_{5}\bar{s}_{7}}{\bar{s}_{3}\bar{s}_{4}-\bar{s}_{7}^{2}},\label{eq:maximizer phi 1-1-1}
\end{equation}
\begin{equation}
\begin{aligned}\left(\tilde{\sigma}\left(\bar{\mathbf{s}}\right)\right)^{2}= & \frac{1}{T-1}\biggl(\bar{s}_{2}+\left(\tilde{\varphi_{0}}\left(\bar{\mathbf{s}}\right)\right)^{2}\bar{s}_{3}+\left(\tilde{\varphi_{1}}\left(\bar{\mathbf{s}}\right)\right)^{2}\bar{s}_{4}-2\tilde{\varphi_{0}}\left(\bar{\mathbf{s}}\right)\bar{s}_{5}\\
 & \hspace{4em}-2\tilde{\varphi_{0}}\left(\bar{\mathbf{s}}\right)\bar{s}_{6}+2\tilde{\varphi_{0}}\left(\bar{\mathbf{s}}\right)\tilde{\varphi_{1}}\left(\bar{\mathbf{s}}\right)\bar{s}_{7}\biggr),
\end{aligned}
\label{eq:maximzer sigma-1}
\end{equation}
and 
\begin{equation}
\tilde{\nu}\left(\bar{\mathbf{s}}\right)=\underset{\nu^{-}<\nu<\nu^{+}}{\arg\max}\ f\left(\nu,\bar{s}_{1}\right),\label{eq:maximizer nu-1}
\end{equation}
where $\bar{s}_{i}$ $\left(i=1,\ldots7\right)$ is the $i$-th component
of $\bar{\mathbf{s}}$. It can be easily verified that $\tilde{\varphi_{0}}\left(\bar{\mathbf{s}}\right)$,
$\tilde{\varphi_{1}}\left(\bar{\mathbf{s}}\right)$ and $\left(\tilde{\sigma}\left(\bar{\mathbf{s}}\right)\right)^{2}$
are continuous functions of $\bar{\mathbf{s}}$, and are 7 times differentiable
with respect to $\bar{\mathbf{s}}$. For $\tilde{\nu}\left(\bar{\mathbf{s}}\right)$,
the gradient of $f\left(\nu,\bar{s}_{1}\right)$ at $\tilde{\nu}$
\begin{equation}
\begin{aligned}g\left(\tilde{\nu},\bar{s}_{1}\right)= & \frac{\partial f\left(\nu,\bar{s}_{1}\right)}{\partial\nu}\Biggl|_{\nu=\tilde{\nu}}\\
= & \frac{1}{2}\left(\log\left(\frac{\tilde{\nu}}{2}\right)-\varPsi\left(\frac{\tilde{\nu}}{2}\right)+1+\frac{\bar{s}_{1}}{T-1}\right)\\
= & 0.
\end{aligned}
\end{equation}
According to the implicit function theorem \cite{Krantz2013}, since
$g\left(\tilde{\nu},\bar{s}_{1}\right)$ is 7 times continuously differentiable
and $\frac{\partial g\left(\tilde{\nu},\bar{s}_{1}\right)}{\partial\tilde{\nu}}=\frac{1}{2}\left(\frac{1}{\tilde{\nu}}-\frac{1}{2}\Psi'\left(\frac{\tilde{\nu}}{2}\right)\right)\neq0$
for any $\tilde{\nu}$ and $\bar{s}_{1}$\cite{liu1995ml}, $\tilde{\nu}\left(\mathbf{s}\right)$
is 7 times continuously differentiable with respect to $\bar{\mathbf{s}}.$

\vspace{-3mm}

\subsection{\textmd{Proof of (SAEM2) and (SAEM3)}}

The condition (SAEM2) has been verified in the proof of the conditions
(M4) and (M5). The condition (SAEM3.1) holds due to the compactness
assumption of the chain in the theorem. The functions $\mathbf{s}\left(\mathbf{y}_{\mathsf{o}},\mathbf{y}_{\mathsf{m}},\boldsymbol{\tau}\right)$
and $\left\{ \hat{\mathbf{s}}^{\left(k\right)}\right\} $ are continuous
function of the chain, therefore, they also take values in a compact
set according to the boundness theorem, which implies the condition
(SAEM3.2) hold.{} Now we focus on the proof of the conditions (SAEM3.3)
and (SAEM3.4).

From the definition of the transition probability $\Pi_{\boldsymbol{\theta}}\left(\mathbf{y}_{\mathsf{m}},\boldsymbol{\tau},\mathbf{y}_{\mathsf{m}}',\boldsymbol{\tau}'\right)$
in \eqref{eq:transition probability}, we can easily verify that the
transition probability $\Pi_{\boldsymbol{\theta}}\left(\mathbf{y}_{\mathsf{m}},\boldsymbol{\tau},\mathbf{y}_{\mathsf{m}}',\boldsymbol{\tau}'\right)$
is continuously differentiable with respect to $\boldsymbol{\theta}$.
In addition, since the derivative is a continuous function of $\boldsymbol{\theta}\in V$
and $\left(\mathbf{y}_{\mathsf{m}},\boldsymbol{\tau},\mathbf{y}_{\mathsf{m}}',\boldsymbol{\tau}'\right)\in\Omega^{2}$,
where $V$ and $\Omega^{2}$ are compact set, according to the boundness
theorem, the derivative is bounded. Therefore, $\Pi_{\boldsymbol{\theta}}\left(\mathbf{y}_{\mathsf{m}},\boldsymbol{\tau},\mathbf{y}_{\mathsf{m}}',\boldsymbol{\tau}'\right)$
is Lipschitz continuous, i.e., for any $\left(\mathbf{y}_{\mathsf{m}},\boldsymbol{\tau},\mathbf{y}_{\mathsf{m}}',\boldsymbol{\tau}'\right)\in\Omega^{2}$,
there exists a real constant $K\left(\mathbf{y}_{\mathsf{m}},\boldsymbol{\tau},\mathbf{y}_{\mathsf{m}}',\boldsymbol{\tau}'\right)$
such that for any $\left(\boldsymbol{\theta},\boldsymbol{\theta}'\right)\in V^{2},$
\begin{equation}
\begin{aligned} & \Bigl|\Pi_{\boldsymbol{\theta}}\left(\mathbf{y}_{\mathsf{m}},\boldsymbol{\tau},\mathbf{y}_{\mathsf{m}}',\boldsymbol{\tau}'\right)-\Pi_{\boldsymbol{\theta}'}\left(\mathbf{y}_{\mathsf{m}},\boldsymbol{\tau},\mathbf{y}_{\mathsf{m}}',\boldsymbol{\tau}'\right)\Bigl|\\
 & \leq K\left(\mathbf{y}_{\mathsf{m}},\boldsymbol{\tau},\mathbf{y}_{\mathsf{m}}',\boldsymbol{\tau}'\right)|\boldsymbol{\theta}-\boldsymbol{\theta}'|.
\end{aligned}
\end{equation}
It follows that 
\begin{equation}
\begin{aligned} & \underset{\left(\mathbf{y}_{\mathsf{m}},\boldsymbol{\tau},\mathbf{y}_{\mathsf{m}}',\boldsymbol{\tau}'\right)\in\Omega^{2}}{\sup}\Bigl|\Pi_{\boldsymbol{\theta}}\left(\mathbf{y}_{\mathsf{m}},\boldsymbol{\tau},\mathbf{y}_{\mathsf{m}}',\boldsymbol{\tau}'\right)-\Pi_{\boldsymbol{\theta}'}\left(\mathbf{y}_{\mathsf{m}},\boldsymbol{\tau},\mathbf{y}_{\mathsf{m}}',\boldsymbol{\tau}'\right)\Bigl|\\
 & \leq L|\boldsymbol{\theta}-\boldsymbol{\theta}'|
\end{aligned}
\end{equation}
with $L=\underset{\left(\mathbf{y}_{\mathsf{m}},\boldsymbol{\tau},\mathbf{y}_{\mathsf{m}}',\boldsymbol{\tau}'\right)\in\Omega^{2}}{\max}K\left(\mathbf{y}_{\mathsf{m}},\boldsymbol{\tau},\mathbf{y}_{\mathsf{m}}',\boldsymbol{\tau}'\right)$,
which implies that the condition (SAEM3.3) is verified.

The condition (SAEM3.4) is about the uniform ergodicity of the Markov
chain generated by the transition probability $\Pi_{\boldsymbol{\theta}}\left(\mathbf{y}_{\mathsf{m}},\boldsymbol{\tau},\thinspace\mathbf{y}_{\mathsf{m}}',\boldsymbol{\tau}'\right)$.
According to Theorem 8 in \cite{roberts2004general}, a Markov chain
is uniformly ergodic, if the transition probability satisfies some
minorization condition, i.e., there exists $\alpha\in N^{+}$ and
some probability measure $\delta\left(\cdot\right)$ such that $\Pi_{\boldsymbol{\theta}}^{\alpha}\left(\mathbf{y}_{\mathsf{m}},\boldsymbol{\tau},\mathbf{y}_{\mathsf{m}}',\boldsymbol{\tau}'\right)\geq\epsilon\delta\left(\mathbf{y}_{\mathsf{m}}',\boldsymbol{\tau}'\right)$
for any $\left(\mathbf{y}_{\mathsf{m}},\boldsymbol{\tau},\mathbf{y}_{\mathsf{m}}',\boldsymbol{\tau}'\right)\in\Omega^{2}$.
Recall our transition probability $\Pi_{\boldsymbol{\theta}}\left(\mathbf{y}_{\mathsf{m}},\boldsymbol{\tau},\thinspace\mathbf{y}_{\mathsf{m}}',\boldsymbol{\tau}'\right)$
is a continuous function for $\left(\mathbf{y}_{\mathsf{m}},\boldsymbol{\tau}\right)\in\Omega$,
according to the extreme value theorem, there must exist an infimum
$g\left(\mathbf{y}_{\mathsf{m}}',\boldsymbol{\tau}',\boldsymbol{\theta}\right)=\underset{\left(\mathbf{y}_{\mathsf{m}},\boldsymbol{\tau}\right)\in\Omega}{\inf}\Pi_{\boldsymbol{\theta}}\left(\mathbf{y}_{\mathsf{m}},\boldsymbol{\tau},\thinspace\mathbf{y}_{\mathsf{m}}',\boldsymbol{\tau}'\right).$
It follows that 
\begin{equation}
\Pi_{\boldsymbol{\theta}}\left(\mathbf{y}_{\mathsf{m}},\boldsymbol{\tau},\thinspace\mathbf{y}_{\mathsf{m}}',\boldsymbol{\tau}'\right)\geq\epsilon\delta\left(\mathbf{y}_{\mathsf{m}}',\boldsymbol{\tau}'\right)
\end{equation}
with $\epsilon=\iint g\left(\mathbf{y}_{\mathsf{m}}',\boldsymbol{\tau}',\boldsymbol{\theta}\right)\mathsf{d}\boldsymbol{\tau}'\mathsf{d}\mathbf{y}_{\mathsf{m}}'$,
and $\delta\left(\mathbf{y}_{\mathsf{m}}',\boldsymbol{\tau}'\right)=\epsilon^{-1}g\left(\mathbf{y}_{\mathsf{m}}',\boldsymbol{\tau}',\boldsymbol{\theta}\right)$.
Therefore, the minorization condition holds in our case, and thus,
the Markov chain generated by $\Pi_{\boldsymbol{\theta}}\left(\mathbf{y}_{\mathsf{m}},\boldsymbol{\tau},\thinspace\mathbf{y}_{\mathsf{m}}',\boldsymbol{\tau}'\right)$
is uniformly ergodic. The condition (SAEM3.4) is verified.

\vspace{-3mm}

 \bibliographystyle{IEEEtran}
\bibliography{../ref}

\end{document}


\title{Supplementary Material}
\author{Junyan~Liu,~ Sandeep~Kumar,~and~Daniel~P.~Palomar}

\maketitle
In this supplementary material, we give detailed proof for the Lemmas
3 and 4:
\begin{description}
\item [{\emph{Lemma\,3\label{lemma3}}}] \emph{For any $\mathbf{y}_{\mathsf{o}}$
and }$\boldsymbol{\theta}\in\Theta$, $p\left(\mathbf{y}_{\mathsf{o}};\boldsymbol{\theta}\right)=\int\int p\left(\mathbf{y},\boldsymbol{\tau};\boldsymbol{\theta}\right)\mathsf{d}\mathbf{y}_{\mathsf{m}}\mathsf{d\boldsymbol{\tau}}=\int p\left(\mathbf{y};\boldsymbol{\theta}\right)\mathsf{d}\mathbf{y}_{\mathsf{m}}<\infty.$ 
\end{description}
\begin{description}
\item [{\emph{Lemma\,4\label{Lemma4}}}] \emph{For any $\mathbf{y}_{\mathsf{o}}$
, $\boldsymbol{\theta}\in\Theta$ and $1<t\leq T$ 
\begin{equation}
\iint g\left(\mathbf{y},\boldsymbol{\tau}\right)p\left(\mathbf{y},\boldsymbol{\tau};\boldsymbol{\theta}\right)\mathsf{d}\mathbf{y}_{\mathsf{m}}\mathsf{d\boldsymbol{\tau}}<\infty,\label{eq:lemma 5}
\end{equation}
where $g\left(\mathbf{y},\boldsymbol{\tau}\right)$ can be $\tau_{t}$,
$\tau_{t}^{2}$, $y_{t}^{2}$,$\tau_{t}y_{t-1}^{2}$, $\tau_{t}y_{t}^{2}$,
or }$-\log\left(\tau_{t}\right)$.
\end{description}
To establish these lemmas, we first introduce some equations and inequalities
in the first section. They are the key ingredients for the proof.
Then we establish Lemma 3 and Lemma 4 in the second and third sections,
respectively. For simplicity of notations, we use $f_{t}\left(y_{j};y_{j-1}\right)$
to denote $f_{t}\left(y_{j};\varphi_{0}+\varphi_{1}y_{j-1},\sigma^{2},\nu\right)$,
$f_{N}\left(y_{j};y_{j-1},\tau_{j}\right)$ to denote $f_{N}\left(y_{j};\varphi_{0}+\varphi_{1}y_{j-1},\frac{\sigma^{2}}{\tau_{j}}\right)$,
and $f_{g}\left(\tau_{j}\right)$ to denote $f_{g}\left(\tau_{j};\frac{\nu}{2},\frac{\nu}{2}\right).$ 

\section{Ingredients}

Recall that, given $\varphi_{0}$, $\varphi_{1},$ $\sigma^{2}$,
$\nu$, and $y_{j-1}$, the variable $y_{j}$ follows a Student's
$t$-distribution: $y_{j}\sim t\left(\varphi_{0}+\varphi_{1}y_{j-1},\sigma^{2},\nu\right)$.
Based on properties of the Student's $t$-distribution, we can get
the following equations and inequality about the variable $y_{j}$
\cite{ahsanullah2014normal}:
\begin{enumerate}
\item The integral of the pdf should be 1:
\begin{equation}
\int f_{t}\left(y_{j};y_{j-1}\right)\mathsf{d}y_{j}=1.\label{eq:integral of pdf}
\end{equation}
\item The first raw moment can be expressed as
\begin{equation}
\int y_{j}f_{t}\left(y_{j};y_{j-1}\right)\mathsf{d}y_{j}=\varphi_{0}+\varphi_{1}y_{j-1}.\label{eq:first moment}
\end{equation}
\item The second raw moment can be expressed as
\begin{equation}
\int y_{j}^{2}f_{t}\left(y_{j};y_{j-1}\right)\mathsf{d}y_{j}=\frac{\nu\sigma^{2}}{\nu-2}+\left(\varphi_{0}+\varphi_{1}y_{j-1}\right)^{2}.\label{eq:second moment}
\end{equation}
\item Since the Student's $t$-distribution can be represented as a Gaussian
mixture \cite{liu1997ml}, the pdf can be rewritten as
\begin{equation}
f_{t}\left(y_{j};y_{j-1}\right)=\int f_{g}\left(\tau_{j}\right)f_{N}\left(y_{j};y_{j-1},\tau_{j}\right)\mathsf{d}\tau_{j}.\label{eq:gaussian mixture integral}
\end{equation}
\item For $\nu>\nu^{-}\geq2$, the pdf of $y_{j}$ can be bounded as
\begin{equation}
f_{t}\left(y_{j};y_{j-1}\right)=\frac{\Gamma\left(\frac{\nu+1}{2}\right)}{\sqrt{\nu\pi}\sigma\Gamma\left(\frac{\nu}{2}\right)}\left(1+\frac{\left(y_{t}-\varphi_{0}-\varphi_{1}y_{t-1}\right)^{2}}{\nu\sigma^{2}}\right)^{-\frac{\nu+1}{2}}\leq\frac{\Gamma\left(\frac{\nu+1}{2}\right)}{\sqrt{\nu\pi}\sigma\Gamma\left(\frac{\nu}{2}\right)}<\infty.\label{eq:pdf upper bound}
\end{equation}
\end{enumerate}
Then we introduce two important inequalities about $\tau_{j}$, which
we will use later. 
\begin{enumerate}
\item The first is about the expectation of $\tau_{j}^{b}$ with $b=1,2$:
\begin{subequations}
\begin{align}
 & \int\tau_{j}^{b}f_{g}\left(\tau_{j}\right)f_{N}\left(y_{j};y_{j-1},\tau_{j}\right)\mathsf{d}\tau_{j}\nonumber \\
 & =\int\frac{\left(\frac{\nu}{2}\right)^{\frac{\nu}{2}}}{\Gamma\left(\frac{\nu}{2}\right)\sqrt{2\pi\sigma^{2}}}\tau_{t}^{\frac{\nu+2b-1}{2}}\exp\left(-\left(\frac{\left(y_{t}-\varphi_{0}-\varphi_{1}y_{t-1}\right)^{2}}{2\sigma^{2}}+\frac{\nu}{2}\right)\tau_{t}\right)\mathsf{d}\tau_{j}\label{eq:taub1}\\
 & =\frac{\left(\frac{\nu}{2}\right)^{\frac{\nu}{2}}}{\Gamma\left(\frac{\nu}{2}\right)\sqrt{2\pi\sigma^{2}}}\frac{\Gamma\left(\frac{\nu+2b+1}{2}\right)}{\left(\frac{\left(y_{j}-\varphi_{0}-\varphi_{1}y_{j-1}\right)^{2}}{2\sigma^{2}}+\frac{\nu}{2}\right)^{\frac{\nu+2b+1}{2}}}\label{eq:taub2}\\
 & =f_{t}\left(y_{j};y_{j-1}\right)\frac{\Gamma\left(\frac{\nu+2b+1}{2}\right)}{\Gamma\left(\frac{\nu+1}{2}\right)\left(\frac{\left(y_{j}-\varphi_{0}-\varphi_{1}y_{j-1}\right)^{2}}{2\sigma^{2}}+\frac{\nu}{2}\right)^{b}}\label{eq:taub3}\\
 & \leq f_{t}\left(y_{j};y_{j-1}\right)\frac{2^{b}\Gamma\left(\frac{\nu+2b+1}{2}\right)}{\nu^{b}\Gamma\left(\frac{\nu+1}{2}\right)},\label{eq:upper bound for tau}
\end{align}
\end{subequations}where the equations (\ref{eq:taub1}) and (\ref{eq:taub3})
follow from the definition of these pdf's, the equation (\ref{eq:taub2})
follows from $\int\frac{\beta^{\alpha}}{\Gamma\left(\alpha\right)}x^{\alpha-1}\exp\left(-\beta x\right)\mathsf{d}x=1$(the
integral of the pdf of the gamma distribution is $1$), the last inequality
(\ref{eq:upper bound for tau}) follows from $\left(\frac{\left(y_{j}-\varphi_{0}-\varphi_{1}y_{j-1}\right)^{2}}{2\sigma^{2}}+\frac{\nu}{2}\right)^{b}\geq\left(\frac{\nu}{2}\right)^{b}.$ 
\item The second inequality is about the expectation of $\log\left(\tau_{j}\right)$,
\begin{subequations}
\begin{align}
 & \int\log\left(\tau_{j}\right)f_{g}\left(\tau_{j}\right)f_{N}\left(y_{j};y_{j-1},\tau_{j}\right)\mathsf{d}\tau_{j}\nonumber \\
 & =\int\frac{\left(\frac{\nu}{2}\right)^{\frac{\nu}{2}}}{\Gamma\left(\frac{\nu}{2}\right)\sqrt{2\pi\sigma^{2}}}\log\left(\tau_{j}\right)\tau_{t}^{\frac{\nu-1}{2}}\exp\left(-\left(\frac{\left(y_{t}-\varphi_{0}-\varphi_{1}y_{t-1}\right)^{2}}{2\sigma^{2}}+\frac{\nu}{2}\right)\tau_{t}\right)\mathsf{d}\tau_{j}\label{eq:logtau1}\\
 & =\frac{\left(\frac{\nu}{2}\right)^{\frac{\nu}{2}}}{\Gamma\left(\frac{\nu}{2}\right)\sqrt{2\pi\sigma^{2}}}\frac{\Gamma\left(\frac{\nu+1}{2}\right)}{\left(\frac{\left(y_{j}-\varphi_{0}-\varphi_{1}y_{j-1}\right)^{2}}{2\sigma^{2}}+\frac{\nu}{2}\right)^{\frac{\nu+1}{2}}}\left(\varPsi\left(\frac{\nu+1}{2}\right)-\log\left(\frac{\left(y_{j}-\varphi_{0}-\varphi_{1}y_{j-1}\right)^{2}}{2\sigma^{2}}+\frac{\nu}{2}\right)\right)\label{eq:logtau2}\\
 & =\left(\varPsi\left(\frac{\nu+1}{2}\right)-\log\left(\frac{\left(y_{j}-\varphi_{0}-\varphi_{1}y_{j-1}\right)^{2}}{2\sigma^{2}}+\frac{\nu}{2}\right)\right)f_{t}\left(y_{j};y_{j-1}\right)\label{eq:logtau3}\\
 & \geq\left(\varPsi\left(\frac{\nu+1}{2}\right)-\frac{\left(y_{j}-\varphi_{0}-\varphi_{1}y_{j-1}\right)^{2}}{2\sigma^{2}}-\frac{\nu}{2}\right)f_{t}\left(y_{j};y_{j-1}\right)\label{eq:logtau4}\\
 & \geq\left(\varPsi\left(\frac{\nu+1}{2}\right)-\frac{\left(y_{j}-\varphi_{0}\right)^{2}+\varphi_{1}^{2}y_{j-1}^{2}}{\sigma^{2}}-\frac{\nu}{2}\right)f_{t}\left(y_{j};y_{j-1}\right)\label{eq:logtau5}\\
 & \geq\left(\varPsi\left(\frac{\nu+1}{2}\right)-\frac{2y_{j}^{2}+2\varphi_{0}^{2}+\varphi_{1}^{2}y_{j-1}^{2}}{\sigma^{2}}-\frac{\nu}{2}\right)f_{t}\left(y_{j};y_{j-1}\right),\label{eq:upper bound for log tau}
\end{align}
\end{subequations}where the equations (\ref{eq:logtau1}) and (\ref{eq:logtau3})
follow from the definition of these pdf's, the equation (\ref{eq:logtau2})
follows from $\int\log\left(x\right)x^{\alpha-1}\exp\left(-\beta x\right)\mathsf{d}x=\frac{\Gamma\left(\alpha\right)}{\beta^{\alpha}}\left(\varPsi\left(\alpha\right)-\log\left(\beta\right)\right)$
\cite{logtau}, the inequality (\ref{eq:logtau4}) follows from $-\log\left(x\right)\geq-x$,
the inequalities (\ref{eq:logtau5}) and (\ref{eq:upper bound for log tau})
follow from $\left(x_{1}+x_{2}\right)^{2}\leq2x_{1}^{2}+2x_{2}^{2}.$
\end{enumerate}

\section{Proof for Lemma 3}

Lemma 3 is about the boundedness of the marginal pdf $p\left(\mathbf{y}_{\mathsf{o}};\boldsymbol{\theta}\right)$.
The pdf can be written as 
\begin{equation}
\begin{aligned}p\left(\mathbf{y}_{\mathsf{o}};\boldsymbol{\theta}\right)= & \int p\left(\mathbf{y};\boldsymbol{\theta}\right)\mathsf{d}\mathbf{y}_{\mathsf{m}}\\
= & \int\prod_{j=2}^{T}f_{t}\left(y_{j};y_{j-1}\right)\mathsf{d}\mathbf{y}_{\mathsf{m}}\\
= & \left(\prod_{d=0}^{D}\prod_{j=t_{d}+n_{d}+2}^{t_{d+1}}f_{t}\left(y_{j};y_{j-1}\right)\right)\left(\prod_{d=1}^{D}\int\prod_{j=t_{d}+1}^{t_{d}+n_{d}+1}f_{t}\left(y_{j};y_{j-1}\right)\mathsf{d}\mathbf{y}_{d}\right),
\end{aligned}
\end{equation}
where we move the term that does not involve $\mathbf{y}_{\mathsf{m}}$
outside the integral.

Since the pdf of the Student's $t$-distribution $f_{t}\left(y_{j};y_{j-1}\right)$
is positive and bounded, we can get the first term $c_{1}\left(\mathbf{y}_{\mathsf{o}},\boldsymbol{\theta}\right)=\prod_{d=0}^{D}\prod_{j=t_{d}+n_{d}+2}^{t_{d+1}}f_{t}\left(y_{j};y_{j-1}\right)<\infty$.
Thus, in order to establish Lemma 3, it is sufficient to establish
the boundedness of the second term, i.e., 
\begin{equation}
\int\prod_{j=t_{d}+1}^{t_{d}+n_{d}+1}f_{t}\left(y_{j};y_{j-1},\sigma^{2},\nu\right)\mathsf{d}\mathbf{y}_{d}<\infty.\label{eq:second term bounded}
\end{equation}

Before carrying out a general proof for the above inequality (\ref{eq:second term bounded}),
we demonstrate the schematic and intuition with a simple example.
We consider a time series as follows: $y_{1},y_{2},\mathsf{NA},\mathsf{NA},\mathsf{NA},y_{6},y_{7}$.
The corresponding second term (\ref{eq:second term bounded}) in this
example can be expressed as \begin{subequations}
\begin{align}
\int\prod_{j=3}^{6}f_{t}\left(y_{j};y_{j-1}\right)\mathsf{d}\mathbf{y}_{1}= & \int\left\{ \int\left(\int f_{t}\left(y_{6};y_{5}\right)f_{t}\left(y_{5};y_{4}\right)\mathsf{d}y_{5}\right)f_{t}\left(y_{4};y_{3}\right)\mathsf{d}y_{4}\right\} f_{t}\left(y_{3};y_{2}\right)\mathsf{d}y_{3}\label{eq:mis block1}\\
\leq & \int\left\{ \int\left(\int\frac{\Gamma\left(\frac{\nu+1}{2}\right)}{\sqrt{\nu\pi}\sigma\Gamma\left(\frac{\nu}{2}\right)}f_{t}\left(y_{5};y_{4}\right)\mathsf{d}y_{5}\right)f_{t}\left(y_{4};y_{3}\right)\mathsf{d}y_{4}\right\} f_{t}\left(y_{3};y_{2}\right)\mathsf{d}y_{3}\label{eqmis block2}\\
= & \int\left\{ \int\frac{\Gamma\left(\frac{\nu+1}{2}\right)}{\sqrt{\nu\pi}\sigma\Gamma\left(\frac{\nu}{2}\right)}f_{t}\left(y_{4};y_{3}\right)\mathsf{d}y_{4}\right\} f_{t}\left(y_{3};y_{2}\right)\mathsf{d}y_{3}\label{eq:mis block3}\\
= & \int\frac{\Gamma\left(\frac{\nu+1}{2}\right)}{\sqrt{\nu\pi}\sigma\Gamma\left(\frac{\nu}{2}\right)}f_{t}\left(y_{3};y_{2}\right)\mathsf{d}y_{3}\label{eq:mis block4}\\
= & \frac{\Gamma\left(\frac{\nu+1}{2}\right)}{\sqrt{\nu\pi}\sigma\Gamma\left(\frac{\nu}{2}\right)}\label{eq:mis block5}\\
< & \infty,\label{eq:mis block 6}
\end{align}
\end{subequations}where the inequality (\ref{eqmis block2}) follows
from (\ref{eq:pdf upper bound}), the equations (\ref{eq:mis block3})-(\ref{eq:mis block5})
hold from (\ref{eq:integral of pdf}), and the inequality (\ref{eq:mis block 6})
follows from the boundness theorem. By applying (\ref{eq:pdf upper bound}),
we find a upper bound $\frac{\Gamma\left(\frac{\nu+1}{2}\right)}{\sqrt{\nu\pi}\sigma\Gamma\left(\frac{\nu}{2}\right)}$
for $f_{t}\left(y_{6};y_{5}\right)$, which does not involve $y_{5}$.
Then the integral about $y_{5}$, $\int\frac{\Gamma\left(\frac{\nu+1}{2}\right)}{\sqrt{\nu\pi}\sigma\Gamma\left(\frac{\nu}{2}\right)}f_{t}\left(y_{5};y_{4}\right)\mathsf{d}y_{5}$,
is easy to compute since $\int f_{t}\left(y_{5};y_{4}\right)\mathsf{d}y_{5}=1$
from (\ref{eq:integral of pdf}), and the result is a function $\frac{\Gamma\left(\frac{\nu+1}{2}\right)}{\sqrt{\nu\pi}\sigma\Gamma\left(\frac{\nu}{2}\right)}$,
which does not involve $y_{4}$. Next, the integral about $y_{4}$,
$\int\frac{\Gamma\left(\frac{\nu+1}{2}\right)}{\sqrt{\nu\pi}\sigma\Gamma\left(\frac{\nu}{2}\right)}f_{t}\left(y_{4};y_{3}\right)\mathsf{d}y_{4}$,
also becomes simple, and the result does not involve $y_{3}$. Finally,
we can get that the integral about $y_{3}$ equals to a continuous
function $\frac{\Gamma\left(\frac{\nu+1}{2}\right)}{\sqrt{\nu\pi}\sigma\Gamma\left(\frac{\nu}{2}\right)}$.
According to the boundness theorem, a continuous function on a closed
bounded set is bounded \cite{royden1988real}, therefore, $\frac{\Gamma\left(\frac{\nu+1}{2}\right)}{\sqrt{\nu\pi}\sigma\Gamma\left(\frac{\nu}{2}\right)}$
is bounded.

Now we come to the general proof for the inequality (\ref{eq:second term bounded}).
The idea is the same as in the example:\begin{subequations}
\begin{align}
\int\prod_{j=t_{d}+1}^{t_{d}+n_{d}+1}f_{t}\left(y_{j};y_{j-1}\right)\mathsf{d}\mathbf{y}_{d}= & \int\cdots\int\int f_{t}\left(y_{t_{d}+n_{d}+1};y_{t_{d}+n_{d}}\right)f_{t}\left(y_{t_{d}+n_{d}};y_{t_{d}+n_{d}-1}\right)\mathsf{d}y_{t_{d}+n_{d}}\nonumber \\
 & \hspace{3em}\left(y_{t_{d}+n_{d}-1};y_{t_{d}+n_{d}-2}\right)\mathsf{d}y_{t_{d}+n_{d}-1}\ldots f_{t}\left(y_{t_{d}+1};y_{t_{d}}\right)\mathsf{d}y_{t_{d}+1}\label{eq:y_d integral 1}\\
\leq & \idotsint\left\{ \int\frac{\Gamma\left(\frac{\nu+1}{2}\right)}{\sqrt{\nu\pi}\sigma\Gamma\left(\frac{\nu}{2}\right)}f_{t}\left(y_{t_{d}+n_{d}};y_{t_{d}+n_{d}-1}\right)\mathsf{d}y_{t_{d}+n_{d}}\right\} \nonumber \\
 & \hspace{3em}\left(y_{t_{d}+n_{d}-1};y_{t_{d}+n_{d}-2}\right)\mathsf{d}y_{t_{d}+n_{d}-1}\ldots f_{t}\left(y_{t_{d}+1};y_{t_{d}}\right)\mathsf{d}y_{t_{d}+1}\label{eq:y_d integral 2}\\
= & \idotsint\frac{\Gamma\left(\frac{\nu+1}{2}\right)}{\sqrt{\nu\pi}\sigma\Gamma\left(\frac{\nu}{2}\right)}\left(y_{t_{d}+n_{d}-1};y_{t_{d}+n_{d}-2}\right)\mathsf{d}y_{t_{d}+n_{d}-1}\ldots f_{t}\left(y_{t_{d}+1};y_{t_{d}}\right)\mathsf{d}y_{t_{d}+1}\label{eq:y_d integral 3}\\
= & \frac{\Gamma\left(\frac{\nu+1}{2}\right)}{\sqrt{\nu\pi}\sigma\Gamma\left(\frac{\nu}{2}\right)}\label{eq:y_d integral 4}\\
< & \infty,\label{eq:y_d integral 5}
\end{align}
\end{subequations}where the inequality (\ref{eq:y_d integral 2})
follows from (\ref{eq:pdf upper bound}), the equations (\ref{eq:y_d integral 2})
and (\ref{eq:y_d integral 3}) hold from (\ref{eq:integral of pdf}),
and the inequality (\ref{eq:y_d integral 5}) follows from the boundness
theorem. Therefore, $p\left(\mathbf{y}_{\mathsf{o}};\boldsymbol{\theta}\right)<\infty.$
Lemma 3 is proved.

\section{Proof for Lemma 4}

Lemma 4 is about the boundedness of  the expectation of $g\left(\mathbf{y},\boldsymbol{\tau}\right)$.
For convenience of proof, we divide the different cases of $g\left(\mathbf{y},\boldsymbol{\tau}\right)$
into four groups:(1) $g\left(\mathbf{y},\boldsymbol{\tau}\right)=\tau_{t}$
or $\tau_{t}^{2}$, (2) $g\left(\mathbf{y},\boldsymbol{\tau}\right)=y_{t}^{2}$,
(3) $g\left(\mathbf{y},\boldsymbol{\tau}\right)$=$\tau_{t}y_{t-1}^{2}$
or $\tau_{t}y_{t}^{2}$, and (4) $-\log\left(\tau_{t}\right)$. We
will prove that the inequality (\ref{eq:lemma 5}) is satisfied for
these groups one by one. 

\subsection{$g\left(\mathbf{y},\boldsymbol{\tau}\right)=\tau_{t}$ or $\tau_{t}^{2}$ }

For $g\left(\mathbf{y},\boldsymbol{\tau}\right)=\tau_{t}^{b}$ with
$b=1,2$, we have\begin{subequations}
\begin{align}
 & \iint g\left(\mathbf{y},\boldsymbol{\tau}\right)p\left(\mathbf{y},\boldsymbol{\tau};\boldsymbol{\theta}\right)\mathsf{d}\mathbf{y}_{\mathsf{m}}\mathsf{d\boldsymbol{\tau}}\nonumber \\
 & =\iint\tau_{t}^{b}p\left(\boldsymbol{\tau};\boldsymbol{\theta}\right)p\left(\mathbf{y}|\boldsymbol{\tau};\boldsymbol{\theta}\right)\mathsf{d}\mathbf{y}_{\mathsf{m}}\mathsf{d\boldsymbol{\tau}}\\
 & =\iint\tau_{t}^{b}\prod_{j=2}^{T}\left\{ f_{g}\left(\tau_{j}\right)f_{N}\left(y_{j};y_{j-1},\tau_{j}\right)\right\} \mathsf{d}\mathbf{y}_{\mathsf{m}}\mathsf{d\boldsymbol{\tau}}\\
 & =\int\Biggl\{\int\tau_{t}^{b}f_{g}\left(\tau_{t}\right)f_{N}\left(y_{t};y_{t-1},\tau_{j}\right)\mathsf{d}\tau_{t}\prod_{j\neq t}\left(\int f_{g}\left(\tau_{j}\right)f_{N}\left(y_{j};y_{j-1},\tau_{j}\right)\mathsf{d}\tau_{j}\right)\Biggr\}\mathsf{d}\mathbf{y}_{\mathsf{m}}\label{eq: tau decompostition}\\
 & \leq\int\frac{2\Gamma\left(\frac{\nu+2b+1}{2}\right)}{\nu\Gamma\left(\frac{\nu+1}{2}\right)}f_{t}\left(y_{t};y_{t-1}\right)\prod_{j\neq t}f_{t}\left(y_{t};y_{t-1}\right)\mathsf{d}\mathbf{y}_{\mathsf{m}}\label{eq:tau inequality}\\
 & =\frac{2^{b}\Gamma\left(\frac{\nu+2b+1}{2}\right)}{\nu^{b}\Gamma\left(\frac{\nu+1}{2}\right)}p\left(\mathbf{y}_{\mathsf{o}};\boldsymbol{\theta}\right)\\
 & <\infty.\label{eq:tau final}
\end{align}
\end{subequations}In (\ref{eq: tau decompostition}), we split the
integral of $\left\{ \tau_{j}\right\} $ into two parts: the first
part involves $\tau_{t}$, while the second does not. In (\ref{eq:tau inequality}),
we apply the inequality (\ref{eq:upper bound for tau}) to the first
part of (\ref{eq: tau decompostition}). The inequality (\ref{eq:tau final})
follows from Lemma 3 and the boundness theorem.

\subsection{$g\left(\mathbf{y},\boldsymbol{\tau}\right)=y_{t}^{2}$}

For $g\left(\mathbf{y},\boldsymbol{\tau}\right)=y_{t}^{2}$, we need
to consider two different cases: $y_{t}$ is observed, and $y_{t}$
is missing. If $y_{t}$ is observed, then we can easily get
\begin{align}
 & \iint g\left(\mathbf{y},\boldsymbol{\tau}\right)p\left(\mathbf{y},\boldsymbol{\tau};\boldsymbol{\theta}\right)\mathsf{d}\mathbf{y}_{\mathsf{m}}\mathsf{d\boldsymbol{\tau}}\nonumber \\
 & =\iint y_{t}^{2}p\left(\mathbf{y},\boldsymbol{\tau};\boldsymbol{\theta}\right)\mathsf{d}\mathbf{y}_{\mathsf{m}}\mathsf{d\boldsymbol{\tau}}\nonumber \\
 & =\int y_{t}^{2}\left\{ \int p\left(\mathbf{y},\boldsymbol{\tau};\boldsymbol{\theta}\right)\mathsf{d\boldsymbol{\tau}}\right\} \mathsf{d}\mathbf{y}_{\mathsf{m}}\nonumber \\
 & =y_{t}^{2}\int p\left(\mathbf{y};\boldsymbol{\theta}\right)\mathsf{d}\mathbf{y}_{\mathsf{m}}\nonumber \\
 & =y_{t}^{2}p\left(\mathbf{y}_{\mathsf{o}};\boldsymbol{\theta}\right)\nonumber \\
 & <\infty.
\end{align}
where the last inequality holds from Lemma 3 and the fact that $\mathbf{y}_{o}$
is finite.

If $y_{t}$ is missing, assume that $y_{t}$ is in the $d_{1}$-th
missing block with $t=t_{d_{1}}+i$, we have\begin{subequations}
\begin{align}
 & \iint g\left(\mathbf{y},\boldsymbol{\tau}\right)p\left(\mathbf{y},\boldsymbol{\tau};\boldsymbol{\theta}\right)\mathsf{d}\mathbf{y}_{\mathsf{m}}\mathsf{d\boldsymbol{\tau}}\nonumber \\
 & =\iint y_{t}^{2}p\left(\mathbf{y},\boldsymbol{\tau};\boldsymbol{\theta}\right)\mathsf{d}\mathbf{y}_{\mathsf{m}}\mathsf{d\boldsymbol{\tau}}\\
 & =\int y_{t}^{2}\left\{ \int p\left(\mathbf{y},\boldsymbol{\tau};\boldsymbol{\theta}\right)\mathsf{d\boldsymbol{\tau}}\right\} \mathsf{d}\mathbf{y}_{\mathsf{m}}\\
 & =\int y_{t}^{2}p\left(\mathbf{y};\boldsymbol{\theta}\right)\mathsf{d}\mathbf{y}_{\mathsf{m}}\\
 & =\prod_{d=0}^{D}\prod_{j=t_{d}+n_{d}+2}^{t_{d+1}}f_{t}\left(y_{j};y_{j-1}\right)\int y_{t}^{2}\prod_{d=1}^{D}\prod_{j=t_{d}+1}^{t_{d}+n_{d}+1}f_{t}\left(y_{j};y_{j-1}\right)\mathsf{d}\mathbf{y}_{d}\label{eq:y decomposition}\\
 & =\prod_{d=0}^{D}\prod_{j=t_{d}+n_{d}+2}^{t_{d+1}}f_{t}\left(y_{j};y_{j-1}\right)\left\{ \prod_{d\neq d_{1}}\int\prod_{j=t_{d}+1}^{t_{d}+n_{d}+1}f_{t}\left(y_{j};y_{j-1}\right)\mathsf{d}\mathbf{y}_{d}\right\} \left\{ \int y_{t_{d_{1}}+i}^{2}\prod_{j=t_{d_{1}}+1}^{t_{d_{1}}+n_{d_{1}}+1}f_{t}\left(y_{j};y_{j-1}\right)\mathsf{d}\mathbf{y}_{d_{1}}\right\} .\label{eq:y decomposition2}
\end{align}
\end{subequations}In (\ref{eq:y decomposition}), we move the term
that does not involve $\mathbf{y}_{\mathsf{m}}=\left\{ \mathbf{y}_{d}\right\} $
outside the integral. In (\ref{eq:y decomposition2}), we split the
integral of $\left\{ \mathbf{y}_{d}\right\} $ into two parts: the
first part does not involve $y_{t_{d_{1}}+i}$, while the second part
does.

Since the pdf $f_{t}\left(y_{j};y_{j-1}\right)$ is bounded from (\ref{eq:pdf upper bound}),
the first term of (\ref{eq:y decomposition2}) is bounded: 
\begin{equation}
\prod_{d=0}^{D}\prod_{j=t_{d}+n_{d}+2}^{t_{d+1}}f_{t}\left(y_{j};y_{j-1}\right)<\infty.
\end{equation}
 In addition, from (\ref{eq:y_d integral 4}), the second term is
also bounded: 
\begin{equation}
\prod_{d\neq d_{1}}\int\prod_{j=t_{d}+1}^{t_{d}+n_{d}+1}f_{t}\left(y_{j};y_{j-1}\right)\mathsf{d}\mathbf{y}_{d}<\infty.
\end{equation}
Thus, in order to prove that (\ref{eq:y decomposition2}) is bounded,
it is sufficient to establish the boundedness of the third term, i.e.,
\begin{equation}
\int y_{t_{d_{1}}+i}^{2}\prod_{j=t_{d_{1}}+1}^{t_{d_{1}}+n_{d_{1}}+1}f_{t}\left(y_{j};y_{j-1}\right)\mathsf{d}\mathbf{y}_{d_{1}}<\infty.\label{eq:boundness of the third term}
\end{equation}

Before carrying out a general proof for the inequality (\ref{eq:boundness of the third term}),
we demonstrate the schematic and intuition with a simple example.
Again, we consider the time series as follows: $y_{1},y_{2},\mathsf{NA},\mathsf{NA},\mathsf{NA},y_{6},y_{7}$.
Then we have\begin{subequations}
\begin{align}
 & \int y_{4}^{2}\prod_{j=3}^{6}f_{t}\left(y_{j};y_{j-1}\right)\mathsf{d}\mathbf{y}_{1}\nonumber \\
 & =\int\left\{ \int\left\{ \int f_{t}\left(y_{6};y_{5}\right)f_{t}\left(y_{5};y_{4}\right)\mathsf{d}y_{5}\right\} y_{4}^{2}f_{t}\left(y_{4};y_{3}\right)\mathsf{d}y_{4}\right\} f_{t}\left(y_{3};y_{2}\right)\mathsf{d}y_{3}\\
 & \leq\int\left\{ \int\left(\int\frac{\Gamma\left(\frac{\nu+1}{2}\right)}{\sqrt{\nu\pi}\sigma\Gamma\left(\frac{\nu}{2}\right)}f_{t}\left(y_{5};y_{4}\right)\mathsf{d}y_{5}\right)y_{4}^{2}f_{t}\left(y_{4};y_{3}\right)\mathsf{d}y_{4}\right\} f_{t}\left(y_{3};y_{2}\right)\mathsf{d}y_{3}\label{eq:y1}\\
 & =\int\left\{ \int\frac{\Gamma\left(\frac{\nu+1}{2}\right)}{\sqrt{\nu\pi}\sigma\Gamma\left(\frac{\nu}{2}\right)}y_{4}^{2}f_{t}\left(y_{4};y_{3}\right)\mathsf{d}y_{4}\right\} f_{t}\left(y_{3};y_{2}\right)\mathsf{d}y_{3}\label{eq:y2}\\
 & =\int\frac{\Gamma\left(\frac{\nu+1}{2}\right)}{\sqrt{\nu\pi}\sigma\Gamma\left(\frac{\nu}{2}\right)}\left(\frac{\nu\sigma^{2}}{\nu-2}+\left(\varphi_{0}+\varphi_{1}y_{3}\right)^{2}\right)f_{t}\left(y_{3};y_{2}\right)\mathsf{d}y_{3}\label{eq:y3}\\
 & =\int\frac{\Gamma\left(\frac{\nu+1}{2}\right)}{\sqrt{\nu\pi}\sigma\Gamma\left(\frac{\nu}{2}\right)}\left(\frac{\nu\sigma^{2}}{\nu-2}+\varphi_{0}^{2}+2\varphi_{0}\varphi_{1}y_{3}+\varphi_{1}^{2}y_{3}^{2}\right)f_{t}\left(y_{3};y_{2}\right)\mathsf{d}y_{3}\label{eq:y4}\\
 & =\frac{\Gamma\left(\frac{\nu+1}{2}\right)}{\sqrt{\nu\pi}\sigma\Gamma\left(\frac{\nu}{2}\right)}\left(\frac{\nu\sigma^{2}}{\nu-2}+\varphi_{0}^{2}\right)+\frac{\Gamma\left(\frac{\nu+1}{2}\right)}{\sqrt{\nu\pi}\sigma\Gamma\left(\frac{\nu}{2}\right)}2\varphi_{0}\varphi_{1}\int y_{3}f_{t}\left(y_{3};y_{2}\right)\mathsf{d}y_{3}+\frac{\Gamma\left(\frac{\nu+1}{2}\right)}{\sqrt{\nu\pi}\sigma\Gamma\left(\frac{\nu}{2}\right)}\varphi_{1}^{2}\int y_{3}^{2}f_{t}\left(y_{3};y_{2}\right)\mathsf{d}y_{3}\label{eq:y5}\\
 & =\frac{\Gamma\left(\frac{\nu+1}{2}\right)}{\sqrt{\nu\pi}\sigma\Gamma\left(\frac{\nu}{2}\right)}\left(\frac{\nu\sigma^{2}}{\nu-2}+\varphi_{0}^{2}\right)+\frac{\Gamma\left(\frac{\nu+1}{2}\right)}{\sqrt{\nu\pi}\sigma\Gamma\left(\frac{\nu}{2}\right)}2\varphi_{0}\varphi_{1}\left(\varphi_{0}+\varphi_{1}y_{2}\right)+\frac{\Gamma\left(\frac{\nu+1}{2}\right)}{\sqrt{\nu\pi}\sigma\Gamma\left(\frac{\nu}{2}\right)}\varphi_{1}^{2}\left(\frac{\nu\sigma^{2}}{\nu-2}+\left(\varphi_{0}+\varphi_{1}y_{2}\right)^{2}\right)\label{eq:y6}\\
 & <\infty,\label{eq:y7}
\end{align}
\end{subequations}where the inequality (\ref{eq:y1}) follows from
(\ref{eq:pdf upper bound}), the equations (\ref{eq:y2})-(\ref{eq:y6})
follow from (\ref{eq:integral of pdf})-(\ref{eq:second moment}),
and the inequality (\ref{eq:y7}) holds due to the boundness theorem.
By applying (\ref{eq:pdf upper bound}), we find an upper bound $\frac{\Gamma\left(\frac{\nu+1}{2}\right)}{\sqrt{\nu\pi}\sigma\Gamma\left(\frac{\nu}{2}\right)}$
for $f_{t}\left(y_{6};y_{5}\right)$, which does not involve $y_{5},$
so that the integral about $y_{5},$ $\int\frac{\Gamma\left(\frac{\nu+1}{2}\right)}{\sqrt{\nu\pi}\sigma\Gamma\left(\frac{\nu}{2}\right)}f_{t}\left(y_{5};y_{4}\right)\mathsf{d}y_{5}$,
is simple and equals to $\frac{\Gamma\left(\frac{\nu+1}{2}\right)}{\sqrt{\nu\pi}\sigma\Gamma\left(\frac{\nu}{2}\right)}$,
which does not involve $y_{4}$. Then the following integrals about
$y_{4}$ and $y_{3}$ are also simple and the final result is bounded.

Now we come to the general proof for the inequality (\ref{eq:boundness of the third term}).
The idea is the same as in the above example:\begin{subequations}
\begin{align}
 & \int y_{t_{d_{1}}+i}^{2}\prod_{j=t_{d_{1}}+1}^{t_{d_{1}}+n_{d_{1}}+1}f_{t}\left(y_{j};y_{j-1}\right)\mathsf{d}\mathbf{y}_{d_{1}}\nonumber \\
 & =\int\cdots\idotsint\biggl\{\int f_{t}\left(y_{t_{d_{1}}+n_{d_{1}}+1};y_{t_{d_{1}}+n_{d_{1}}}\right)f_{t}\left(y_{t_{d_{1}}+n_{d_{1}}};y_{t_{d_{1}}+n_{d_{1}}-1}\right)\mathsf{d}y_{t_{d_{1}}+n_{d_{1}}}\biggr\}\hspace{6.5em}\nonumber \\
 & \hspace{7.2em}f_{t}\left(y_{t_{d_{1}}+n_{d_{1}}-1};y_{t_{d_{1}}+n_{d_{1}}-2}\right)\mathsf{d}y_{t_{d_{1}}+n_{d_{1}}-1}\ldots\nonumber \\
 & \hspace{7.2em}y_{t_{d_{1}}+i}^{2}f_{t}\left(y_{t_{d_{1}}+i};y_{t_{d_{1}}+i-1}\right)\mathsf{d}y_{t_{d_{1}}+i}\ldots f_{t}\left(y_{t_{d_{1}}+1};y_{t_{d_{1}}}\right)\mathsf{d}y_{t_{d_{1}}+1}\label{eq:yg1}\\
 & \leq\int\cdots\idotsint\biggl\{\int\frac{\Gamma\left(\frac{\nu+1}{2}\right)}{\sqrt{\nu\pi}\sigma\Gamma\left(\frac{\nu}{2}\right)}f_{t}\left(y_{t_{d_{1}}+n_{d_{1}}};y_{t_{d_{1}}+n_{d_{1}}-1}\right)\mathsf{d}y_{t_{d_{1}}+n_{d_{1}}}\biggr\}\nonumber \\
 & \hspace{7.2em}f_{t}\left(y_{t_{d_{1}}+n_{d_{1}}-1};y_{t_{d_{1}}+n_{d_{1}}-2}\right)\mathsf{d}y_{t_{d_{1}}+n_{d_{1}}-1}\ldots\nonumber \\
 & \hspace{7.2em}y_{t_{d_{1}}+i}^{2}f_{t}\left(y_{t_{d_{1}}+i};y_{t_{d_{1}}+i-1}\right)\mathsf{d}y_{t_{d_{1}}+i}\ldots f_{t}\left(y_{t_{d_{1}}+1};y_{t_{d_{1}}}\right)\mathsf{d}y_{t_{d_{1}}+1}\label{eq:yg2}\\
 & =\int\cdots\idotsint\frac{\Gamma\left(\frac{\nu+1}{2}\right)}{\sqrt{\nu\pi}\sigma\Gamma\left(\frac{\nu}{2}\right)}f_{t}\left(y_{t_{d_{1}}+n_{d_{1}}-1};y_{t_{d_{1}}+n_{d_{1}}-2}\right)\mathsf{d}y_{t_{d_{1}}+n_{d_{1}}-1}\ldots\nonumber \\
 & \hspace{4.5em}y_{t_{d_{1}}+i}^{2}f_{t}\left(y_{t_{d_{1}}+i};y_{t_{d_{1}}+i-1}\right)\mathsf{d}y_{t_{d_{1}}+i}\ldots f_{t}\left(y_{t_{d_{1}}+1};y_{t_{d_{1}}}\right)\mathsf{d}y_{t_{d_{1}}+1}\hspace{6.5em}\label{eq:yg3}
\end{align}
\begin{align}
 & =\idotsint\frac{\Gamma\left(\frac{\nu+1}{2}\right)}{\sqrt{\nu\pi}\sigma\Gamma\left(\frac{\nu}{2}\right)}y_{t_{d_{1}}+i}^{2}f_{t}\left(y_{t_{d_{1}}+i};y_{t_{d_{1}}+i-1}\right)\mathsf{d}y_{t_{d_{1}}+i}\ldots f_{t}\left(y_{t_{d_{1}}+1};y_{t_{d_{1}}}\right)\mathsf{d}y_{t_{d_{1}}+1}\label{eq:yg4}\\
 & =\idotsint\frac{\Gamma\left(\frac{\nu+1}{2}\right)}{\sqrt{\nu\pi}\sigma\Gamma\left(\frac{\nu}{2}\right)}\left(\frac{\nu\sigma^{2}}{\nu-2}+\left(\varphi_{0}+\varphi_{1}y_{t_{d_{1}}+i-1}\right)^{2}\right)\mathsf{d}y_{t_{d_{1}}+i-1}\ldots f_{t}\left(y_{t_{d_{1}}+1};y_{t_{d_{1}}}\right)\mathsf{d}y_{t_{d_{1}}+1}\label{eq:yg5}
\end{align}
\end{subequations}where the inequality (\ref{eq:yg2}) follows from
(\ref{eq:pdf upper bound}), the equations (\ref{eq:yg3}) and (\ref{eq:yg4})
hold from (\ref{eq:integral of pdf}), and the equation follows from
(\ref{eq:second moment}). Similar to the example, this integral of
(\ref{eq:yg5}) will finally reduce to a quadratic function of $y_{t_{d_{1}}}$.
Then, according to the boundness theorem, we can obtain
\begin{equation}
\int y_{t_{d_{1}}+i}^{2}\prod_{j=t_{d_{1}}+1}^{t_{d_{1}}+n_{d_{1}}+1}f_{t}\left(y_{j};y_{j-1}\right)\mathsf{d}\mathbf{y}_{d_{1}}<\infty,
\end{equation}
and thus, $\iint y_{t}^{2}p\left(\mathbf{y},\boldsymbol{\tau};\boldsymbol{\theta}\right)\mathsf{d}\mathbf{y}_{m}\mathsf{d\boldsymbol{\tau}}<\infty$.

\subsection{$g\left(\mathbf{y},\boldsymbol{\tau}\right)$=$\tau_{t}y_{t-1}^{2}$
or \textmd{\normalsize{}$\tau_{t}y_{t}^{2}$} }

In this subsection, we consider the cases of $g\left(\mathbf{y},\boldsymbol{\tau}\right)$=$\tau_{t}y_{t}^{2}$
or $\tau_{t}y_{t-1}^{2}$. Here we only present proof for the case
of $g\left(\mathbf{y},\boldsymbol{\tau}\right)=\tau_{t}y_{t}^{2}$,
the case $\tau_{t}y_{t-1}^{2}$ can be verified similarly. For $g\left(\mathbf{y},\boldsymbol{\tau}\right)=\tau_{t}y_{t}^{2}$,
we have\begin{subequations}
\begin{align}
 & \iint g\left(\mathbf{y},\boldsymbol{\tau}\right)p\left(\mathbf{y},\boldsymbol{\tau};\boldsymbol{\theta}\right)\mathsf{d}\mathbf{y}_{\mathsf{m}}\mathsf{d\boldsymbol{\tau}}\nonumber \\
 & =\iint\tau_{t}y_{t}^{2}p\left(\mathbf{y},\boldsymbol{\tau};\boldsymbol{\theta}\right)\mathsf{d}\mathbf{y}_{\mathsf{m}}\mathsf{d\boldsymbol{\tau}}\\
 & =\iint\tau_{t}y_{t}^{2}\prod_{j=2}^{T}\left\{ f_{N}\left(y_{j};y_{j-1},\tau_{j}\right)f_{g}\left(\tau_{j}\right)\right\} \mathsf{d}\mathbf{y}_{\mathsf{m}}\mathsf{d\boldsymbol{\tau}}\\
 & =\int y_{t}^{2}\Biggl\{\int\tau_{t}f_{N}\left(y_{t};y_{t-1},\tau_{t}\right)f_{g}\left(\tau_{t}\right)\mathsf{d}\tau_{t}\prod_{j\neq t}\left(\int f_{N}\left(y_{j};y_{j-1},\tau_{j}\right)f_{g}\left(\tau_{j}\right)\mathsf{d}\tau_{j}\right)\Biggr\}\mathsf{d}\mathbf{y}_{\mathsf{m}}\label{eq:tau_y1}\\
 & \leq\int y_{t}^{2}\frac{2\Gamma\left(\frac{\nu+3}{2}\right)}{\nu\Gamma\left(\frac{\nu+1}{2}\right)}f_{t}\left(y_{t};y_{t-1}\right)\prod_{j\neq t}f_{t}\left(y_{j};y_{j-1}\right)\mathsf{d}\mathbf{y}_{\mathsf{m}}\label{eq:tau_y2}\\
 & =\frac{2\Gamma\left(\frac{\nu+3}{2}\right)}{\nu\Gamma\left(\frac{\nu+1}{2}\right)}\int y_{t}^{2}p\left(\mathbf{y}_{\mathsf{o}},\mathbf{y}_{\mathsf{m}};\boldsymbol{\theta}\right)\mathsf{d}\mathbf{y}_{\mathsf{m}}\label{eq:tau_y3}\\
 & <\infty.\label{eq:tau_y4}
\end{align}
\end{subequations}In (\ref{eq:tau_y1}), we split the integral of
$\boldsymbol{\tau}=\left\{ \tau_{j}\right\} $ into two parts: the
first part involves $\tau_{t}$, while the second does not. The inequality
(\ref{eq:tau_y2}) holds from (\ref{eq:upper bound for tau}). The
inequality (\ref{eq:tau_y4}) follows from the result of last subsection
and the boundness theorem.

\subsection{\textmd{\normalsize{}$g\left(\mathbf{y},\boldsymbol{\tau}\right)=-\log\left(\tau_{t}\right)$} }

Finally, we consider the case of $g\left(\mathbf{y},\boldsymbol{\tau}\right)=-\log\left(\tau_{t}\right)$:\begin{subequations}
\begin{align}
 & \iint g\left(\mathbf{y},\boldsymbol{\tau}\right)p\left(\mathbf{y},\boldsymbol{\tau};\boldsymbol{\theta}\right)\mathsf{d}\mathbf{y}_{\mathsf{m}}\mathsf{d\boldsymbol{\tau}}\nonumber \\
 & =-\iint\log\left(\tau_{t}\right)p\left(\mathbf{y},\boldsymbol{\tau};\boldsymbol{\theta}\right)\mathsf{d}\mathbf{y}_{\mathsf{m}}\mathsf{d\boldsymbol{\tau}}\\
 & =-\iint\log\left(\tau_{t}\right)\prod_{j=2}^{T}\left\{ f_{N}\left(y_{j};y_{j-1},\tau_{j}\right)f_{g}\left(\tau_{j}\right)\right\} \mathsf{d}\mathbf{y}_{\mathsf{m}}\mathsf{d\boldsymbol{\tau}}\\
 & =-\int\Biggl\{\int\log\left(\tau_{t}\right)f_{N}\left(y_{t};y_{t-1},\tau_{t}\right)f_{g}\left(\tau_{t}\right)\mathsf{d}\tau_{t}\prod_{j\neq t}\left(\int f_{N}\left(y_{j};y_{j-1},\tau_{j}\right)f_{g}\left(\tau_{j}\right)\mathsf{d}\tau_{j}\right)\Biggr\}\mathsf{d}\mathbf{y}_{\mathsf{m}}\hspace{3.5em}\label{eq:tau_y1-1}
\end{align}
\begin{align}
 & \leq-\int\left(\varPsi\left(\frac{\nu+1}{2}\right)-\frac{2y_{t}^{2}+2\varphi_{0}^{2}+\varphi_{1}^{2}y_{t-1}^{2}}{\sigma^{2}}-\frac{\nu}{2}\right)f_{t}\left(y_{t};y_{t-1}\right)\prod_{j\neq t}f_{t}\left(y_{j};y_{j-1}\right)\mathsf{d}\mathbf{y}_{\mathsf{m}}\label{eq:tau_y2-1}\\
 & =\int\left(\frac{2y_{t}^{2}+2\varphi_{0}^{2}+\varphi_{1}^{2}y_{t-1}^{2}}{\sigma^{2}}+\frac{\nu}{2}-\varPsi\left(\frac{\nu+1}{2}\right)\right)p\left(\mathbf{y}_{\mathsf{o}},\mathbf{y}_{\mathsf{m}};\boldsymbol{\theta}\right)\mathsf{d}\mathbf{y}_{\mathsf{m}}\label{eq:tau_y3-1}\\
 & =\frac{2}{\sigma^{2}}\int y_{t}^{2}p\left(\mathbf{y}_{\mathsf{o}},\mathbf{y}_{\mathsf{m}};\boldsymbol{\theta}\right)\mathsf{d}\mathbf{y}_{\mathsf{m}}+\frac{\varphi_{1}^{2}}{\sigma^{2}}\int y_{t-1}^{2}p\left(\mathbf{y}_{\mathsf{o}},\mathbf{y}_{\mathsf{m}};\boldsymbol{\theta}\right)\mathsf{d}\mathbf{y}_{\mathsf{m}}+\left(\frac{2\varphi_{0}^{2}}{\sigma^{2}}+\frac{\nu}{2}-\varPsi\left(\frac{\nu+1}{2}\right)\right)p\left(\mathbf{y}_{\mathsf{o}};\boldsymbol{\theta}\right)\\
 & <\infty.\label{eq:tau_y4-1}
\end{align}
\end{subequations}In (\ref{eq:tau_y1-1}), we split the integral
of $\boldsymbol{\tau}=\left\{ \tau_{j}\right\} $ into two parts:
the first part involves $\tau_{t}$, while the second does not. The
inequality (\ref{eq:tau_y2-1}) holds from (\ref{eq:upper bound for log tau}).
The inequality (\ref{eq:tau_y4-1}) is from the result of the last
subsection, Lemma 3 and the boundness theorem. 

\bibliographystyle{IEEEtran}
\bibliography{../ref}